\documentclass[aps,twocolumn,groupedaddress,showpacs,showkeys,nofootinbib]{revtex4}

\newcommand{\Det}{{\textrm {Det}}}

\newcommand{\D}{{\mathcal D}}

\def\slashchar#1{\setbox0=\hbox{$#1$}
   \dimen0=\wd0 \setbox1=\hbox{/} \dimen1=\wd1
   \ifdim\dimen0>\dimen1 \rlap{\hbox to \dimen0{\hfil/\hfil}} #1
   \else  \rlap{\hbox to \dimen1{\hfil$#1$\hfil}} / \fi}
\def\p{\slashchar{p}}
\def\w{\omega} 
\def\D{{\bf D}}

\begin{document}

\title{Low Energy Chiral Lagrangian in Curved Space-Time from the
Spectral Quark Model}

\author{E. Meg\'{\i}as}
\email{emegias@ugr.es}

\author{E. \surname{Ruiz Arriola}}
\email{earriola@ugr.es}

\author{L. L. Salcedo}
\email{salcedo@ugr.es}

\affiliation{
Departamento de F\'{\i}sica Moderna,
Universidad de Granada,
E-18071 Granada, Spain}

\author{W. Broniowski} 
\email{Wojciech.Broniowski@ifj.edu.pl} 
\affiliation{The H. Niewodnicza\'nski Institute of Nuclear Physics,
Polish Academy of Sciences, PL-31342 Cracow, Poland}

\date{\today} 

\begin{abstract}
We analyze the recently proposed Spectral Quark Model in the light of
Chiral Perturbation Theory in curved space-time. In particular, we
calculate the chiral coefficients $L_1, \dots, L_{10}$, as well as the
coefficients $L_{11}$, $L_{12}$, and $L_{13}$, appearing when the model
is coupled to gravity.  The analysis is carried for the SU(3) case. We
analyze the pattern of chiral symmetry breaking as well as elaborate
on the fulfillment of anomalies.  Matching the model results to
resonance meson exchange yields the relation between the masses of the
scalar, tensor and vector mesons, $M_{f_0}=M_{f_2}=\sqrt{2} M_V= 4
\sqrt{3 /N_c} \pi f_\pi$. Finally, the large-$N_c$ limit suggests the
dual relations in the vector and scalar channels, $M_V=M_S= 2 \sqrt{6
/N_c} \pi f_\pi $ and $\langle r^2 \rangle^{1/2}_S = \langle r^2
\rangle^{1/2}_V = 2 \sqrt{N_c} / f_\pi = 0.59 {\rm fm} $.
\end{abstract}

\pacs{12.38.Lg}

\keywords{quark models; chiral Lagrangean; Chiral Perturbation Theory;
anomalies; curved space-time.}

\maketitle

\section{INTRODUCTION}
\label{sec:intro}

The low-energy structure of QCD in the presence of external
electroweak and gravitational sources is best described by Chiral
Perturbation Theory
(ChPT)~\cite{Langacker:1973hh,Weinberg:1978kz,Gasser:1983yg,Gasser:1984gg,Donoghue:qv}
(for review see, {\em e.g.}, Ref.~\cite{Pich:1995bw}). In the meson
sector, the spontaneous breaking of chiral symmetry dominates at low
energies and systematic calculations of the corresponding low-energy
constants (LEC's) have been carried out in the recent past up to two
loop
accuracy~\cite{Bijnens:1998fm,Colangelo:2001df,Amoros:2001cp,Bijnens:2002hp},
or by using the Roy equations~\cite{Ananthanarayan:2000ht} (see also
~\cite{Yndurain:2002ud,Pelaez:2003eh}). For strong and electroweak
processes involving pseudoscalar mesons the bulk of the LEC's is
saturated in terms of resonance exchanges~\cite{Ecker:1988te}, which
can be justified in the large-$N_c$ limit in a certain low-energy
approximation~\cite{Pich:2002xy} by imposing the QCD short-distance
constraints. In the case of gravitational processes similar ideas
apply~\cite{Donoghue:qv}, although less information is
known~\cite{Kubis:1999db}.  Nowadays, ChPT can be used as a qualitative
and quantitative test to any model of low-energy hadron structure.
 
In the quest to understand the microscopic dynamics underlying the
LEC's, their calculation in chiral quark models has been undertaken
many times~\cite{Diakonov:tw,Balog:ps,Andrianov:ay,Belkov:mb,
Espriu:1989ff,Hansson:jy,Holdom:iq,RuizArriola:gc,Bernard:1991wy,
Schuren:1991sc,Bijnens:1992uz,Schuren:1993aj,Polyakov:1995vh,Wang:1999cp}.
The effort has been made to compute $L_1, \dots L_{10}$, which
correspond to the flat-space-time case. The calculation of
$L_{11},L_{12} $ and $L_{13}$, encoding the coupling to gravitational
sources, has not been considered so far. Roughly speaking, these
calculations are generally described in terms of some long-wavelength
expansion of the fermion determinant associated to the
constituent-quark degrees of freedom. A detailed scrutiny shows,
however, that the implementation of the necessary regularization is
not always satisfactory from several viewpoints. The
regularization of a low-energy chiral quark model
corresponds to a physical suppression of the high-energy quark states.
This can be achieved in a number of different ways, {\em e.g.} by
cut-offs, form factors, or momentum-dependent masses, provided they do
not break symmetries such as the gauge invariance and chiral
symmetry. Thus, the regularization should not be removed in the
end. In such a situation, where the  high-energy quark states are
suppressed above a certain scale $\Lambda$, one should expect a
power-like behaviour $ \Lambda^n / Q^n $ for any large-momentum
external leg of the quark loop in the high-momentum limit. In the
language of the parton model this high-energy behaviour corresponds to
the onset of scaling.

As a matter of fact, one of the questions which could not be answered by  
low-energy calculations 
 concerns the low-energy resolution scale where these models are
supposedly defined. Actually, in order to properly answer this
question one should look instead into {\em high-energy processes} and
demand parton-model relations on the constituent quarks. As pointed
out in Ref.~\cite{RuizArriola:2002wr}, a sensible scheme is obtained
by demanding that the momentum fraction carried by the valence quarks in
a hadron saturates the energy-momentum sum rule. Once this initial
scale is defined one can use the QCD evolution to compute an
observable at a higher scale. This way the QCD radiative corrections
are incorporated. In fact, using the analysis of the Durham group
carried out a decade ago~\cite{Sutton:1991ay} for the case of the
pion, one obtains the result that the valence quarks saturate the
energy-momentum sum rule at $ \mu_0 = 313~{\rm MeV}$ if the LO DGLAP
QCD perturbative analysis is carried out. Although this scale looks
quite low, the impressive agreement obtained for the parton
distribution functions of the pion after the DGLAP evolution in
LO~\cite{Davidson:1994uv,Davidson:2001cc} and
NLO~\cite{Davidson:2001cc} (see also Ref.~\cite{RuizArriola:2002wr},
and Ref.~\cite{Broniowski:2003rp} where the comparison to the E615
data \cite{e615} is made) supports this interpretation of the low
resolution scale. Moreover, using that scale, the pion distribution
amplitude~\cite{RuizArriola:2002bp} and the off-forward generalized
parton functions~\cite{Broniowski:2003rp} agree well also with the
recent transverse lattice
calculations~\cite{Dalley:2002nj,Dalley:2003sz}, which presumably
incorporate the non-perturbative evolution.

A proper identification of the low-energy matrix elements entering the
high-energy processes is grounded on the absence of logarithmic
corrections in the low-energy model in the high-energy limit, since
the proper QCD radiative logarithmic corrections are automatically
and completely
incorporated by the QCD evolution. Not surprisingly, this condition
imposes severe constraints on the kind of admissible regularization
schemes. In a recent work the Spectral Quark Model (SQM) has been
proposed~\cite{RuizArriola:2001rr,RuizArriola:2003bs}, implementing the
so-called spectral regularization (see below) complying to these 
power-like short distance constraints. 

In the present paper we extend the SQM to the SU(3) flavor group and
include finite current quark mass. Instead of using the construction
of vertices based on the Ward-Takahashi identities, employed in
Ref.~\cite{RuizArriola:2001rr,RuizArriola:2003bs}, it is by far more
convenient to define the effective action depending on the non-linear
pseudoscalar meson fields in the presence of external scalar,
pseudoscalar, vector, axial, and gravitational sources. The latter
have never been considered in chiral quark model calculations. This
effective action is defined in Sect.~\ref{sec:eff_ac}. 
 We also show
in Sect.~\ref{sec:moments} how one can explicitly eliminate the
spectral function in terms of the quark momentum-dependent mass and
wave-function renormalization. Following the standard procedure we
perform the gradient expansion of the spectral-regularized fermion
determinant both for the anomalous (Sect.~\ref{sec:anomalies}) as well
as the non-anomalous sectors in curved space-time
(Sect.~\ref{sec:low_ener}). As a consequence the structure of the
energy-momentum tensor may be analyzed. Remarkably, our spectral
regularization method complies to the QCD anomaly without removing the
regularization. Therefore the standard
Wess-Zumino-Witten~\cite{Wess:yu,Witten:tw} term is generated for a
finite regularization. In the non-anomalous sector we find, through
the comparison to the standard chiral
Lagrangian~\cite{Gasser:1983yg,Gasser:1984gg}, that the low-energy
constants at ${\cal O} (p^4)$ associated to terms which are
non-vanishing in the chiral limit are completely independent of the
regularization details. The LEC's associated to terms carrying the
current quark mass coefficients do depend on the particular ansatz for
the spectral regularization, and we evaluate them using the
regularization based on the meson dominance of form
factors~\cite{RuizArriola:2003bs}. Such a model has provided a
satisfactory description of the quark self-energy of the recent
lattice data~\cite{Bowman:2002bm}. Finally (Sect.~\ref{sec:dual}), we
also confront the large-$N_c$ relations~\cite{Pich:2002xy} and 
discuss the consequences of extending the present model to include
these constraints. The Appendix \ref{sec:app} 
 contains details of
the formalism in the
 curved space-time.

\section{THE EFFECTIVE ACTION OF THE SPECTRAL QUARK MODEL} 
\label{sec:eff_ac}

In a recent work the spectral quark model (SQM) has been
introduced~\cite{RuizArriola:2001rr,RuizArriola:2003bs}. The
approach is similar in spirit to the model of Efimov and Ivanov 
\cite{Efimov:1988yd}, proposed many years ago. It is based on the
formal introduction of the generalized Lehmann representation for the
quark propagator,
\begin{eqnarray}
S({p}) = \int_C d \omega { \rho( \omega ) \over
\slashchar{p} - \omega  } \equiv \frac{Z(p^2)}{\p - M(p^2)},
\label{eq:lehmann} 
\end{eqnarray}
where $\rho(\omega)$ is a (generally complex) quark spectral function
and $C$ denotes a suitable contour in the complex $\omega$ plane. 
The function $M(p^2)$ is the quark self-energy, while $Z(p^2)$ is the
quark wave-function renormalization. In the case of {\em analytic
confinement}, {\em i.e.}, when the propagator does not have poles, a
sensible definition of a constituent quark mass is (from now 
on we drop the index $C$ from the $\w$ integral, which is implicitly
understood to run along the contour $C$)  
\begin{eqnarray} 
M_Q = M(0) = \int d \w \frac{\rho ( \w)}{\w} \Big/ \int d \w
\frac{\rho ( \w)}{\w^2} .
\label{eq:q_mass} 
\end{eqnarray} 
As discussed at length in Ref.~\cite{RuizArriola:2003bs}, the proper
normalization and the conditions of finiteness of hadronic observables
are achieved by requesting an infinite set of {\em spectral
conditions} for the moments of the quark spectral function
$\rho(\omega)$, namely
\begin{eqnarray}
\rho_0 && \equiv \int d\omega \rho(\omega ) = 1, \label{rho0} \\
\rho_n && \equiv \int d\omega \omega^n \rho(\omega) = 0, 
 \;\;\; {\rm for} \; n=1,2,3,... \label{rhon}
\end{eqnarray} 
Physical observables are proportional the zeroth and the inverse
moments,
\begin{eqnarray}
\rho_{-k} && \equiv \int d\omega \omega^{-k} \rho(\omega), \;\;\; {\rm
 for} \; k=0,1, 2, 3,... ,\label{rhoinv}
\end{eqnarray} 
as well as to the {\em ``log moments''},
\begin{eqnarray}
&& \rho^\prime_n\equiv \int d\omega \log(\omega^2/\mu^2) \omega^n
\rho(\omega) \nonumber \\ && =\int d\omega \log(\omega^2) \omega^n
\rho(\omega), \;\;\; {\rm for} \; n=1,2,3,4,...
\label{rholog}
\end{eqnarray}
Obviously, when an observable is proportional to the dimensionless zeroth moment,
$\rho_0=1$, the result does not depend explicitly on the
regularization. The spectral conditions (\ref{rhon}) remove the
dependence on the scale $\mu$ in (\ref{rholog}), thus guaranteeing the
absence of any dimensional transmutation. The only exception is the
$0th$-log moment,
\begin{eqnarray}
\rho^\prime_0 (\mu^2) = \int d \omega \log (\omega^2 /\mu^2) \rho(\omega),
\end{eqnarray}  
which does depend on a scale $\mu$ and is {\it not} regularized by the
spectral method (see the discussion below). No standard requirement of
positivity for the spectral strength, $\rho(\omega)$, is made. Unlike
other regularizations, such as the dimensional regularization or 
the $\zeta$-function regularization, the spectral regularization is 
physical in the sense that it provides a high-energy suppression
in one-quark-loop amplitudes and is not removed at the end of the
calculation. It also improves on a Pauli-Villars regularization,
because it complies to the factorization property of correlation
functions, form factors, {\em etc.}, in the high-energy limit, {\em i.e.}, it
guarantees the absence of logarithmic corrections to form factors.
The phenomenological success of the SQM in describing structure
functions of the pion, generalized parton
distributions~\cite{Broniowski:2003rp,Broniowski:2003wn}, and the pion
light-cone wave function~\cite{RuizArriola:2002bp,RuizArriola:2003wi}
suggests that the whole scheme deserves to be thoroughly 
pursued further.

In Ref.~\cite{RuizArriola:2003bs} it was argued that there are a
number of terms in the one-quark-loop effective low-energy chiral
Lagrangean which correspond to taking the infinite-cut-off limit. The
terms with explicit chiral-symmetry breaking do not correspond to this
class. The purpose of this paper is to analyze these terms, which are
specific both to the regularization and the choice of couplings in the
spectral quark Lagrangian. For completeness we also consider the gauge
couplings and gravitational couplings, which allows us a
determination of all low-energy constants in the SU(3) sector in the
SQM approach.

The effective action complying to the solution of the Ward-Takahashi
identities via the gauge technique of Delbourgo and
West~\cite{Delbourgo:1977jc} corresponds in our case to the minimum
substitution prescription for the spectral quark. It yields a quark
fermionic determinant of the form
\begin{eqnarray}
\Gamma[U,s,p,v,a,g] =-{\rm i} N_c \int d \omega \rho(\omega) {\rm Tr} \log
\left( i {\bf D} \right),
\label{eq:eff_ac} 
\end{eqnarray} 
where the Dirac operator is given by 
\begin{eqnarray}
i {\bf D} &=& i\slashchar{d} - \omega U^5 - {\hat m_0} + \left(
\slashchar{v} + \slashchar{a} \gamma_5 - s - i \gamma^5 p \right) \nonumber \\ 
          &=& i D - \omega U^5.
\label{eq:dirac_op} 
\end{eqnarray} 
The derivative $d_\mu $ is frame (local Lorentz) and general-coordinate covariant
and it includes the spin connection (see Appendix \ref{sec:app} for
notation). The symbols $s$, $p$, $v_\mu$, and $a_\mu $ denote the external scalar,
pseudoscalar, vector, and axial flavour sources, respectively, given in
terms of the generator of the flavour SU(3) group, 
\begin{eqnarray} 
s= \sum_{a=0}^{N_F^2-1} s_a \frac{\lambda_a}2, \qquad \dots  
\end{eqnarray}   
with $\lambda_a $ representing the Gell-Mann matrices. The tensor
$g_{\mu\nu} $ is the metric external source representing the coupling
to a gravitational field. The matrix $U^5 = U^{\gamma_5}$, and $ U =
u^2 = e^{ { i} \sqrt{2} \Phi / f } $ is the flavour matrix
representing the pseudoscalar octet of mesons in the non-linear
representation,
\begin{eqnarray} 
	\Phi = \left( \matrix{ \frac{1}{\sqrt{2}} \pi^0 +
	\frac{1}{\sqrt{6}} \eta & \pi^+ & K^+  \cr  \pi^- & -
	\frac{1}{\sqrt{2}} \pi^0 + \frac{1}{\sqrt{6}} \eta & K^0  \cr 
	K^- & \bar{K}^0 & - \frac{2}{\sqrt{6}} \eta }
	\right) .
\end{eqnarray}
The matrix ${\hat m}_0={\rm Diag}(m_u, m_d, m_s)$ is the current quark
mass matrix and $f$ denotes the pion weak-decay constant in the chiral
limit, to be determined later on from the proper normalization
condition of the pseudoscalar fields. For a bilocal (Dirac and flavour
matrix valued) operator $A(x,x')$ one has 
\begin{eqnarray}
{\rm Tr} A = \int d^4 x
\sqrt{-g} \,{\rm tr} \langle A(x,x) \rangle \, , 
\end{eqnarray} 
with ${\rm tr} $ denoting the Dirac
trace and $\langle \, \rangle $ the flavour trace. Moreover, $g={\rm
det} g_{\mu\nu} $ is the determinant of the curved space-time metric.
Finally, in the second line of Eq.~(\ref{eq:dirac_op}) we have
introduced the Dirac operator $ D$ corresponding to the external
fields only. The $U_A (1)$ is taken into account by extending the
matrix to the U(3) sector, $ U \to \bar U = U e^{{\rm i} \eta_8 / (3
f) } $ with $ \det U =1 $, adding the customary term
\begin{eqnarray}
{\cal L} = - \frac{f^2}4 m_{\eta_1}^2 \left\{ \theta - \frac{ i}2
\left[ \log \det U - \log \det U^\dagger \right] \right\}^2 .
\label{eq:singlet} 
\end{eqnarray} 
The Dirac operator given by Eq.~(\ref{eq:dirac_op}) transforms
covariantly under local chiral transformations (see
Appendix~\ref{sec:app}). 

Formally, in the flat space-time the effective action
(\ref{eq:eff_ac}) looks quite familiar and we should point out here
that the main difference with similar actions, such as, {\em e.g.}, the
one of Ref.~\cite{Espriu:1989ff}, is related to the regularization
procedure. Actually, the method of Ref.~\cite{Espriu:1989ff} consists
of taking $\rho(\omega) = \delta (\omega -M_Q) $ with $M_Q$ being the
constituent quark mass. This choice satisfies the normalization
condition $\rho_0=1$, but does not comply to the $\rho_n =0$ spectral
requirements. The problem can be avoided if one uses suitable regularization
methods, such as the dimensional or $\zeta$-function regularization, but
then {\em logarithmic corrections} to form factors are
generated and the well-known Landau instability found long ago in Refs.~\cite{Ripka:am,Soni:fy} sets in. 

The pion form factor obtained from the
$\zeta$-function regularization used for instance in Ref.~\cite{Espriu:1989ff}
for $t=-Q^2 $ becomes, in the chiral limit,
$$ 
F (Q^2) = -\frac{ 4 N_c M_Q^2 }{(4 \pi)^2 f_\pi^2} \int_0^1 dx \log
\left[\frac{ x (1-x) Q^2 + M_Q^2}{\mu^2} \right],
$$  
where the pion weak-decay constant is given by $f_\pi^2 = 4 N_c M_Q^2
\log (\mu^2 / M_Q^2) / (4 \pi)^2 $. While the proper normalization
$F(0)=1$ is obtained, at large momenta one has a logarithmic behavior,
$ F(Q^2) \to \log ( Q^2) $, instead of the power-like behaviour, which
poses a problem. On the other hand, the spectral regularization method
yields~\cite{RuizArriola:2003wi} $ F(Q^2) \to N_c / 4 \pi^2 f_\pi^2 (
2 \rho_4' / Q^2 - 2\rho_6' / Q^4 + \dots )$, with no logs
present. This twist expansion property allows us to extract in a clean
way the low-energy matrix elements relevant for high-energy
processes~\cite{RuizArriola:2002wr}.

Given the fact that the integration contour is in general complex,
passing to the Euclidean space and separating the action into the real and
imaginary parts becomes a bit inconvenient. Instead, we take the full
advantage of the Minkowski space and introduce the auxiliary operator,
\begin{eqnarray}
-{\rm i} {\bf D}_5 &=& \gamma_5 \left( {\rm i} \slashchar{d} -\w {U^5}^\dagger -\hat m_0 +
\slashchar{v} - \gamma_5 \slashchar{a} -
s + i\gamma_5 p \right) \gamma_5, \nonumber \\ 
\end{eqnarray} 
which corresponds to the Hermitean conjugation in the Euclidean space. Thus,
the normal parity action is given by
\begin{eqnarray}
S_{\rm n.p.}= -\frac{\rm i}2 N_c \int d \w \rho(\w) {\rm Tr} \log
\left( {\bf D} {\bf D}_5 \right).
\label{eq:np}
\end{eqnarray} 

\section{Relation of spectral moments to quark mass and normalization}
\label{sec:moments}

A potential disadvantage of the spectral regularization is that the inverse
problem, {\em i.e.} the problem of finding the spectral function
$\rho(\w)$ from the known moments, does not always have an easy explicit solution or
perhaps has no solution at all. In this section we show how the negative
moments and the log-moments can be translated into the integrals involving
the quark mass function, $M(p^2)$, and the quark wave function
renormalization, $Z(p^2)$. Let us start with Eq.~(\ref{eq:lehmann})
and assume that the set of spectral conditions is met
\begin{eqnarray}
\int d \w \w^n \rho(\w) = \delta_{n0}, \;\;\;\; n=0, 1, \dots . 
\end{eqnarray}
Then, the following identity, proved by induction, holds: 
\begin{eqnarray}
\int d \w { \w^n \rho( \w ) \over
\slashchar{p} - \w } = \p^n S(\p) - \p^{n-1} .
\end{eqnarray}
Rationalizing the denominators yields 
\begin{eqnarray}
\int d \w \w^n \rho( \w ) \frac{\slashchar{p} + \w }{p^2-\w^2} = \p^n
Z(p^2) \frac{\p + M(p^2)}{p^2-M(p^2)^2} - \p^{n-1}. \nonumber \\
\end{eqnarray}
We have two cases of odd and even $n$. For $n=2k$ we find  
\begin{eqnarray}
\int \!\!\! d \w \w^{2k} \rho( \w ) \frac{\slashchar{p} + \w }{p^2- \w^2} &=& p^{2k} 
Z(p^2) \frac{\p + M(p^2)}{p^2-M(p^2)^2} - \p p^{2k-2}. \nonumber \\
\label{rela}
\end{eqnarray}
Defining 
\begin{eqnarray}
L_n (p^2 ) =  \int d \w \w^{n} \rho( \w ) \frac1{p^2- \w^2}
\end{eqnarray} 
and comparing coefficients of powers of $\p$ in Eq.~ref{rela} produces
the identities  
\begin{eqnarray}
L_{2k} (p^2) &=& p^{2k} Z(p^2) \frac1{p^2-M(p^2)^2} - p^{2k-2}, \nonumber \\
L_{2k+1}(p^2) &=& p^{2k} Z(p^2) \frac{M(p^2)}{p^2-M(p^2)^2}.
\end{eqnarray}
The case $n=2k+1$ produces the same relations.  

The following recursion relations follow 
directly from the spectral conditions (\ref{rhon}):
\begin{eqnarray}
\int d \w { \w^{n} \rho( \w ) \over p^2 - \w^2 } &=&
p^2 \int d \w { \w^{n-2} \rho( \w ) \over p^2 -
\w^2 }, \;\;\;n>2, \nonumber \\
\end{eqnarray} 
which are obvious when on the right-hand side we write $p^2=(p^2-\w^2)+\w^2$.
We now pass to the Euclidean space, $\p^2 = p^2 \to -p_E^2 $, 
and get
\begin{eqnarray} 
\int d \w \w^n \log(\w^2) \rho(\w) &=& \int_0^\infty dp_E^2 L_n (-
p_E^2 ) \label{era1} \\&=& -\int_0^\infty dp_E^2 p_E^2 L_{n-2} (- p_E^2 ). \nonumber
\end{eqnarray}
Thus, we have obtained the log-moments in terms of $Z$ and $M$. The negative
moments are simply derivatives of the quark propagator at the origin,
\begin{eqnarray}
\int d \w \frac{\rho(\w)}{\w^n} = -\left( \frac{d}{d\p} \right)^{n-1}
S(\p) \Big|_{p=0} \qquad n=1,2, \dots . \label{era2}
\end{eqnarray} 
The derivative is computed taking $p^2 = \p \p $. Thus, given the
quark propagator $S(\p)$ we may just use formulas
(\ref{era1},\ref{era2}) to translate negative moments and log-moments
without ever having to specify explicitly the spectral function. This
is a rather remarkable feature of the spectral approach. The
expressions for $f_\pi$, $\langle \bar q q \rangle $ (the quark
condensate for a single flavor), and $B$ (the vacuum energy density)
in the chiral limit are
\begin{eqnarray}
f^2 &=& \frac{4 N_c}{(4\pi)^2} \int d \w \w^2 \rho( \w) (-\log \w^2),
\\ \langle \bar q q \rangle &=& \frac{4 N_c}{(4\pi)^2} \int d \w \w^3
\rho(\w) (-\log \w^2), \\ -B &=& \frac{N_F N_c}{(4\pi)^2} \int d \w \w^4
\rho(\w) (-\log \w^2) =\frac14 \langle \theta^\mu_\mu \rangle,
\label{eq:vac}
\end{eqnarray}  
respectively. Here, $\theta^{\mu\nu}$ is the energy momentum tensor
(see also Sect.~\ref{sec:ener-mom}). We get, for instance,
\begin{eqnarray}
f^2 =  \frac{ 4 N_c}{(4\pi)^2} \int_0^\infty dp_E^2
\frac{M(-p_E^2)^2- p_E^2 (Z(-p_E^2)-1)}{p_E^2+ M(-p_E^2)^2},
\label{fpins}
\end{eqnarray}
or
\begin{eqnarray}
\langle \bar q q \rangle =  \frac{ 4 N_c}{(4\pi)^2} \int_0^\infty dp_E^2 p_E^2
\frac{Z(-p_E^2)M(-p_E^2)}{p_E^2+ M(-p_E^2)^2}.
\label{qqns}
\end{eqnarray}
In Eq.~(\ref{qqns}) we recognize the usual formula for the quark
condensate found in non-local models. On the other hand,
Eq.~(\ref{fpins}) is different from analogous quark-model
expressions \cite{Pagels:hd,Bowler:ir}. The reason is that, strictly
speaking, the above formulas should only be used for functions $
M(p^2)$ and $ Z(p^2) $ complying to the generalized Lehmann
representation, Eq.~(\ref{eq:lehmann}), with the spectral density satisfying the spectral
conditions.

One can use similar manipulations to get the pion electromagnetic form factor
obtained in Ref.~\cite{RuizArriola:2003bs}. For space-like
momentum, $Q^2=-q^2$, we obtain
\begin{eqnarray} 
F_V ( Q^2 ) &=& \frac{ 4 N_c}{(4\pi)^2 f_\pi^2} \int_0^1 dx
\\ & \times &\int_0^\infty dp_E^2 \frac{M(-P_E^2 )^2-
P_E^2(Z(-P_E^2)-1)}{P_E^2+ M(-P_E^2)^2}, \nonumber
\label{eq:formfac} 
\end{eqnarray} 
where 
\begin{eqnarray}
P_E^2 = p_E^2 + x(1-x) Q^2 . 
\end{eqnarray} 
Note that the inversion procedure used in
Ref.~\cite{RuizArriola:2003bs} to determine the spectral density from
vector meson dominance (the Meson Dominance version of the SQM) is
linear, whereas written in terms of $M$ and $Z$ becomes highly non-linear. 

\section{Chiral anomalies} 
\label{sec:anomalies}

One of the major advantages of the spectral regularization is that it makes
hadronic observables finite and scale independent, a basic
requirement of any regularization procedure. However, that does not
necessarily mean nor imply that the full effective action in the
presence of external fields is finite, since even in the case of the
vanishing pion fields, $U=1$, we have non-hadronic
processes. Actually, it turns out that the photon wave function
renormalization \cite{RuizArriola:2003bs} is proportional to $\rho_0'$, 
thus it depends on the scale $\mu$ and therefore diverges in
some regularization schemes (such as the dimensional
regularization). This scale dependence arises also in other
non-hadronic terms of the effective action. 

In Ref.~\cite{RuizArriola:2003bs} it was checked that the $\pi^0 \to 2
\gamma $ and $ \gamma \to 3 \pi $ decays comply to the correct
values expected from the chiral QCD anomaly. With the help of the
effective action, Eq.~(\ref{eq:eff_ac}), we now want to show that this is
also true for all anomalous processes. In order to understand the role
of regularization, it is instructive to compute the chiral anomaly
first. Next, we will show that in the presence of external fields the
anomaly does not depend on the pion field $U$ and thus coincides with the
anomaly in QCD due to the spectral conditions
$\rho_1= \rho_2=\rho_3 = \rho_4=0$.

Under chiral (vector and axial) local transformations the
Dirac operator transforms as
\begin{eqnarray} 
 \D \to e^{+i\epsilon_V (x) -i\epsilon_A (x) \gamma_5 } \D
e^{-i\epsilon_V (x)-i\epsilon_A (x) \gamma_5 },
\end{eqnarray} 
with
\begin{eqnarray} 
\epsilon_V (x) = \sum_a  \epsilon_V^a (x) \lambda_a , \qquad
\epsilon_A (x) = \sum_a  \epsilon_A^a (x) \lambda_a.
\end{eqnarray} 
Infinitesimally, we have
\begin{eqnarray}
\delta \D = i [ \epsilon_V , \D] - i \{ \epsilon_A , \D \}  .
\end{eqnarray} 
If we make a chiral transformation of the effective action~(\ref{eq:eff_ac}) 
without any additional regularization, we get
\begin{eqnarray}
\delta S = - i N_c {\rm Tr} \int d \w \rho (\w) \left[ \delta \D
\D^{-1} \right] . 
\end{eqnarray} 
If we assume the cyclic property of the functional trace we get a
contribution from the axial variation only,  
\begin{eqnarray} 
\delta_A S &\equiv&
{\cal A}_A = \int d^4 x \,{\rm tr} \int d \w \rho(\w) \langle 2 i \alpha \gamma_5
\rangle \nonumber \\ &=& \rho_0 \int d^4 x \,{\rm tr} \langle 2 i \alpha \gamma_5 \rangle ,
\end{eqnarray}  
a result which, due to the infinite dimensional
trace~\cite{Fujikawa:1979ay,Fujikawa:1980eg}, is ambiguous even in the presence of
the spectral regularization. Thus, to get rid of the ambiguity we have
to introduce an extra regularization. As is well known, there is no
regularization preserving the chiral symmetry, thus the anomaly is generated.

The calculation can be done by standard methods. A very convenient one
is the $\zeta$-function regularization~\cite{Salcedo:1994qy}, which
computes the anomaly directly in terms of the Dirac operator itself
(and not its square) and does not require any redefinition of the
Dirac $\gamma_5$ matrix. This yields the equation
\begin{eqnarray} 
\delta_A S &\equiv& {\cal A}_A = {\rm Tr} \int d \w \rho(\w) \left( 2 i
\alpha \gamma_5 \left[ {\rm i} {\bf D} \right]^0 \right) \\
&=& \int d^4 x \,{\{\rm tr} \int d \w \rho(\w) \langle 2 i
\alpha (x) \gamma_5 \langle x | {\bf D}^0 | x
\rangle \rangle , \nonumber
\end{eqnarray}  
where the zeroth power of the Dirac operator is understood as an
analytical continuation which can be written in terms of the
Seeley-DeWitt coefficients for the Dirac operators~\cite{Salcedo:1994qy},
\begin{eqnarray}
\langle x | {\bf D}^0 | x \rangle &=& \frac1{(4\pi)^2} \Big\{ {1\over 2}\D^4 +
               {1\over 3}(\D^2\Gamma_\mu^2+\Gamma_\mu\D^2\Gamma_\mu +
               \Gamma_\mu^2\D^2) \nonumber \\ &&+ {1\over 6}
               \left(\Gamma_\mu^2\Gamma_\nu^2 + (\Gamma_\mu\Gamma_\nu)^2 +
               \Gamma_\mu\Gamma_\nu^2\Gamma_\mu \right) \Big\} ,
\end{eqnarray} 
where $\Gamma_\mu = {1\over 2}\{\gamma_\mu,\D\}$ and the operator $\D$
acts to the left. The result for general couplings in four dimensions
has been obtained from Ref.~\cite{Salcedo:1994qy}. Direct inspection
shows that since the $\w$-dependence is given by $i \D = i D - \w U^5
$, the result can be written as a sum of an $\w$-independent term and
a polynomial remainder,
\begin{eqnarray} 
{\cal A}_A &=& \int d \w \rho( \w) \left( {\cal A}_A [ v,a,s,p ] + 
{\cal A}_A [ v,a,s,p, \w,U] \right) \nonumber \\
&=&\rho_0 {\cal A}_A [ v,a,s,p ] ,
\label{eq:qcd_anom} 
\end{eqnarray}  
where the $\w $-dependent polynomial term vanishes due to the spectral
conditions. This shows that the anomaly of the spectral quark mode
coincides with the anomaly of QCD {\it after}
 introducing an
additional suitable regularization, regardless of the details of the
spectral function. This result is common also to nonlocal models when
one evaluates
anomalies~\cite{RuizArriola:1998zi,RuizArriola:1999us}. This is an
important point since if the effective action $ \Gamma [U,s,p,v,a] $
in Eq.~(\ref{eq:eff_ac}) is both chiral symmetric and finite there
is apparently no reason for anomalies. We will see below how and where these
divergences arise.

To see now how the standard
Wess-Zumino-Witten~\cite{Wess:yu,Witten:tw} term arises in the present
context, let us consider for simplicity the chiral limit $ \hat m_0 =
0$ and set the external fields to zero and work in flat space, so that
$ {\rm i} {\bf D} = {\rm i} \slashchar{\partial} $. A convenient
representation can be obtained by introducing the field
\begin{eqnarray} 
U_t^5 = e^{ { i} t \sqrt{2} \gamma_5 \Phi /f},
\end{eqnarray} 
interpolating between the vacuum, $U_{t=0}^5=1$, and the full matrix $U_{t=1}^5 = U^5$. 
Then, we have the trivial but useful identity for the vacuum-subtracted
action, 
\begin{eqnarray}
\Gamma[U,s, \dots ] &-& \Gamma[1,s, \dots ] = \\ -&i& N_c \int_0^1 dt
\frac{d}{dt} \int_C d \omega \rho(\omega) {\rm Tr} \log \left( {\rm i}
D - \omega U^5_t \right) \nonumber \\ = &i& \int_0^1 dt \int_C d
\omega \rho(\omega) {\rm Tr}\left[ \omega \frac{d U_t^5}{dt} \frac1{
{\rm i} D - \omega U^5_t } \right] . \nonumber
\label{eq:GammaU}
\end{eqnarray} 
Using the representation in Eq.~(\ref{eq:GammaU}) and the formulas of
Appendix~\ref{sec:app} the result can be obtained
straightforwardly. Since we are interested in abnormal parity
processes it is enough to identify the terms containing the
Levi-Civita tensor $ \epsilon_{\mu\nu\alpha \beta} $, which due to
the Lorentz invariance requires at least four derivatives. Taking into
account the fact that the derivative operator acts to the right we get
\begin{eqnarray}
S_{\rm ab}^{(4)} &=& - i N_c \int_0^1 dt \int_C d \omega \rho(\omega)
\int d^4 x \int \frac{d^4 k}{(2 \pi)^4 } \frac1{[k^2- \omega^2]^5} \nonumber \\ &\times & {\rm Tr}\left\{ -\omega \gamma_5
U_t^\dagger \frac{d U_t }{dt} \omega \left[ \omega U_t^\dagger {\rm i}
\slashchar{\partial} U_t \right]^4 \right\}.
\end{eqnarray}  
After computation of the traces and integrals we finally find 
\begin{eqnarray}
\Gamma_{\rm ab}^{(4)} &=& \rho_0 \frac{N_c}{48\pi^2} \int_0^1 dt \int d^4 x
\epsilon_{\mu\nu\alpha\beta} \\ &\times& \langle U_t^\dagger
\frac{d U_t }{dt} U_t^\dagger \partial^\mu U_t U_t^\dagger
\partial^\nu U_t U_t^\dagger \partial^\alpha U_t U_t^\dagger
\partial^\beta U_t \rangle , \nonumber 
\label{eq:wzw} 
\end{eqnarray}  
which coincides with the WZW term if the spectral normalization
condition $\rho_0 =1 $ is used. External fields can be included again
through the use of Eq.~(\ref{eq:GammaU}), yielding the gauged WZW
term in the Bardeen subtracted form. Actually, the difference
$\Gamma[U,s,p,v,a]-\Gamma[1,s,p,v,a] $ is finite and preserves gauge
invariance but breaks chiral symmetry generating the anomaly of
Eq.~(\ref{eq:qcd_anom}).

Higher order corrections to the abnormal parity component of the
action involve negative spectral moments. For instance, the terms ${\cal O}
(p^6) $ and higher are regularized, and involve $\rho_{-2}$ for terms
with no quark mass terms and $\rho_{-1}$ for terms containing one
quark mass. This is in contrast to the approach of
Ref.~\cite{Espriu:1989ff} where the infinite cut-off limit is
considered for a constant constituent quark mass. In this
regard let us also note that for the unregularized abnormal parity
action one would get the transition form factor
$$
F_{\pi \gamma \gamma^* } ( Q^2 ) =  \frac{8  M_Q^2 }{ ( 4 \pi)^2
f_\pi} \int_0^1 dx \frac1{(1-x) x Q^2 + 2 M_Q^2 } 
$$ 
which satisfies the proper anomaly condition, $ F_{\pi \gamma \gamma^*
} ( 0 )= 1 /(4 \pi^2 f_\pi)$. Again, a log dependent term is obtained
at high virtualities (see also Ref.~\cite{RuizArriola:2002wr}), in
contrast to the correct twist expansion generated by the spectral
method~\cite{RuizArriola:2003bs}.

\section{Low energy chiral expansion of the action} 
\label{sec:low_ener}

The chiral expansion of the action, Eq.~(\ref{eq:eff_ac}), correponds
to a counting where the pseudoscalar field $U$ and the curved
space-time metric $g^{\mu \nu} $ are zeroth order, the vector and
axial fields $ v_\mu $ and $a_\mu$ are first order, and any derivative
$\partial_\mu $ first order. The external scalar and pseudoscalar
fields $s$ and $p$ and the current mass matrix $\hat m_0$ are taken to
be second order. In chiral quark models at the one loop level this
chiral expansion corresponds to a derivative expansion. With the help
of the action of Eq.~(\ref{eq:eff_ac}) one can compute the derivative
expansion in curved space-time (see Appendix \ref{sec:app} for
details),
\begin{eqnarray}
S= \int d^4 x \sqrt{-g} {\cal L} (x)   
\end{eqnarray}  
where the effective chiral Lagrangian in the Gasser, Leutwyler and
Donoghue form~\cite{Gasser:1984gg,Donoghue:qv} reads
\begin{eqnarray} 
{\cal L} &=& {\cal L}^{(0)}+{\cal L}^{(2,g)} + {\cal L}^{(2,R)} +
{\cal L}^{(4,g)} + {\cal L}^{(4,R)} + \dots, \nonumber \\ 
\label{eq:chl}
\end{eqnarray} 
with the metric (upperscript $g$) and curvature (upperscript $R$)
terms explicitly separated. The zeroth order vacuum contribution reads
\begin{eqnarray} 
{\cal L}^{(0)} = B = \frac{N_F N_c}{(4\pi)^2} \rho_4^\prime \, , 
\label{eq:chi0} 
\end{eqnarray} 
where the vacuum constant is given by Eq.~(\ref{eq:vac}).

\subsection{Metric contributions} 

The metric contributions read
\begin{eqnarray} 
{\cal L}^{(2,g)}  &=& {f^2\over 4} \langle D_\mu U^\dagger D^\mu U
+(\chi^\dagger U + U^\dagger \chi) \rangle ,
\label{eq:chl2}
\end{eqnarray} 
and 
\begin{eqnarray} 
{\cal L}^{(4,g)} &=& L_1 \langle D_\mu U^\dagger D^\mu U \rangle^2 + L_2
  \langle D_\mu U^\dagger D_\nu U \rangle^2  \nonumber \\ &+& L_3
  \langle \left( D_\mu U^\dagger D_\nu U \right)^2\rangle \nonumber \\ &+& L_4
  \langle D_\mu U^\dagger D^\mu U \rangle \langle \chi^\dagger U +
  U^\dagger \chi \rangle \nonumber \\ &+& L_5 \langle D_\mu U^\dagger
  D^\mu U ( \chi^\dagger U + U^\dagger \chi) \rangle \nonumber \\ &+& L_6 \langle
  \chi^\dagger U + U^\dagger \chi \rangle^2 \nonumber \\ &+& L_7
  \langle \chi^\dagger U - U^\dagger \chi \rangle^2 + L_8 \langle (
  \chi^\dagger U)^2 + (U^\dagger \chi)^2 \rangle \nonumber \\ &-& iL_9
  \langle F_{\mu\nu}^L D_\mu U D_\nu U^\dagger + F_{\mu\nu}^R D_\mu
  U^\dagger D_\nu U \rangle \nonumber \\ &+& L_{10} \langle
  F_{\mu\nu}^L U F_{\mu\nu}^R U^\dagger \rangle \nonumber \\ &+& H_1
  \langle (F_{\mu\nu}^R)^2 + (F_{\mu\nu}^L)^2 \rangle + H_2 \langle
  \chi^\dagger \chi \rangle.
\label{eq:chl4}
\end{eqnarray} 
We have introduced the standard chiral covariant derivatives and gauge
field strength tensors,
\begin{eqnarray} 
D_\mu U &=& D_\mu^L U-U D_\mu^R =
\partial_\mu U-i A_\mu^L U +iU  A_\mu^R, \\ 
 F_{\mu\nu}^r &=& i[ D_\mu^r, D_\nu^r] = 
\partial_\mu A_\nu^r -\partial_\nu A_\mu^r
-i [ A_\mu^r , A_\nu^r ], \nonumber 
\end{eqnarray} 
with $r=L, R$.
The pion weak-decay constant and the quark condensate in the chiral limit read
\begin{eqnarray}
f^2 &=& - \frac{ 4 N_c}{(4\pi)^2} \rho_2' , \\ f^2 B_0 &=& -
\langle \bar q q  \rangle =
\frac{4N_c}{(4\pi)^2} \rho_3',
\end{eqnarray} 
while the chiral coefficients are~\footnote{\label{foot}The value of $L_7$ displayed here
corresponds to the SU(3) model only. For the U(3) model one gets
$L_7=0$ but then the term of Eq.~(\ref{eq:singlet}) should be added, and the 
value of $L_7$ is changed.}  
\begin{eqnarray}
L_3 &=& -2 L_2 = -4 L_1 = -\frac{N_c}{(4 \pi)^2} \frac{\rho_0}6,  \\
L_4 &=& L_6=0, \\ L_5 &=& -\frac{N_c}{(4\pi)^2} \frac{\rho_1'}{2 B_0}, \\ 
L_7 &=& \frac{ N_c }{(4 \pi)^2} \frac1{2 N_F}\left(  \frac{\rho_1'}{2B_0}
+ \frac{\rho_0}{12} \right) ,\\
L_8 &=& \frac{N_c}{(4\pi)^2} \left[ \frac{\rho_2'}{ 4 B_0^2 } - 
\frac{\rho_1'}{4 B_0} - \frac{\rho_0}{24} \right] ,\\ 
L_9 &=& -2 L_{10} = \frac{ N_c }{(4 \pi)^2} \frac{\rho_0}3 ,\\ 
H_1 &=& \frac{ N_c }{(4 \pi)^2} \frac{\rho^\prime_0}6  ,\\
H_2 &=& \frac{ N_c }{(4 \pi)^2} \left( \frac{\rho^\prime_2}{B_0^2} +  
\frac{\rho^\prime_1}{2 B_0} + \frac{\rho_0}{12} \right) ,
\label{eq:lec's_g}
\end{eqnarray} 
where $N_F=2,3$. As we can see, the coefficients $L_1, L_2,L_3, L_4,
L_6 , L_9 , L_{10} $ are pure numbers, and coincide for convergent
integrals with those expected in the limit where the regularization is
removed~\cite{Espriu:1989ff}. The argument anticipating this result in
Ref.~\cite{RuizArriola:2003bs} has to do with the dimensionless
character of the low energy couplings which thus involve the zeroth
moment $\rho_0=1$. Note that this remarkable result holds {\em
without} removing the regularization~\footnote{Actually, the kinetic
energy term obtained in Ref.~\cite{Espriu:1989ff} within the
$zeta$-function regularization was scale dependent, so dimensional
transmutation sets in. If dimensional regularization is used, it would
lead to a $1/\epsilon$-divergence, which after renormalization would
also lead to dimensional transmutation. The point of the spectral
regularization is that dimensional transmutation is precluded thanks
to the spectral conditions, Eqs.~(\ref{rhon}), and {\it any} 
choice of the spectral function
yields the
same finite result.}. The fact that $H_1$ is proportional to $\rho_0^\prime
$ corresponds to a scale-dependent or divergent gauge-field wave
function, and was observed already in
Ref.~\cite{RuizArriola:2003bs}. Hence, the finite piece of $H_1$ 
depends on the regularization scheme.   

We can use $f $ and $L_{5}$ in order to
determine $L_7$, $L_8 $, $B_0$  and $H_2$, which immediately yields 
\begin{eqnarray}
L_7&=&-\frac{L_5}{2N_f}+\frac{N_c}{384 \pi^2 N_f} \simeq -0.35 \cdot
10^{-3}, \nonumber \\
 L_8&=&\frac{L_5}{2}-\frac{N_c}{384
\pi^2}-\frac{f^2}{64B_0^2} \simeq 1.05 \cdot 10^{-3}, \nonumber \\
H_2&=&L_5+\frac{N_c}{192 \pi^2}-\frac{f^2}{4B_0^2} \simeq 2.1 \cdot
10^{-3} .
\end{eqnarray}
The numerical values displayed here 
have been obtained with the large-$N_c$ value of $L_5$ from Table I.

\subsection{Curvature contributions} 

The curvature contributions to the chiral Lagrangian can be written in
the form proposed in Ref.~\cite{Donoghue:qv} and are given by 
\begin{eqnarray}
{\cal L}^{(2,R)} &=& H_0 R \, 
\end{eqnarray} 
and 
\begin{eqnarray}
{\cal L}^{(4,R)} &=& -L_{11} R \langle D^\mu U^\dagger D_\mu U \rangle 
-L_{12} R^{\mu \nu} \langle D_\mu U^\dagger D_\nu U \rangle \nonumber
\\ &-& L_{13} R \langle \chi^\dagger U + U^\dagger \chi \rangle + H_3 R^2 + H_4
R_{\mu \nu} R^{\mu \nu} \nonumber \\ &+&  H_5 R_{\mu \nu \alpha \beta} R^{\mu \nu
\alpha \beta}.
\label{eq:chl4R}
\end{eqnarray} 
Here $R^\lambda_{\sigma \mu \nu}$, $R_{\mu \nu} $ and $R$ are the
Riemann curvature tensor, the Ricci tensor, and the curvature scalar,
respectively\footnote{Note the opposite sign of our definition for
the Riemann tensor as compared to Ref.~\cite{Donoghue:qv}. We follow
Ref.~\cite{We-grav} (see Appendix \ref{sec:app}).},
\begin{eqnarray}
-R^\lambda_{\,\, \sigma \mu \nu} &=& \partial_\mu \Gamma^\lambda_{\nu \sigma}
- \partial_\nu \Gamma^\lambda_{\mu \sigma}+ \Gamma^\lambda_{\mu
\alpha} \Gamma^\alpha_{\nu \sigma} - \Gamma^\lambda_{\nu \alpha}
\Gamma^\alpha_{\mu \sigma}, \nonumber \\ 
R_{\mu \nu} &=& R^\lambda_{\, \, \mu \lambda \nu } \, ; \quad R = g^{\mu \nu} R_{\mu
\nu} \, .
\label{eq:curvature}    
\end{eqnarray}
The Christoffel symbols are specified in
Eq.~(\ref{eq:christoffel}). The curvature terms reflect the composite
nature of the pseudoscalar fields, since in the considered model they
correspond to the coupling of the gravitational external field at the
quark level. After some algebra we get
\begin{eqnarray}
H_0 &=& -\frac{f^2}4 \frac1{12}, \\ 
L_{12} &=& - 2 L_{11} = -\frac{N_c}{(4\pi)^2} \frac{\rho_0}6 , 
\\ L_{13} &=& -\frac{N_c}{(4\pi)^2} \frac{\rho_1^\prime}{12 B_0} = \frac16 L_5,  \\ 
H_3 &=& -\frac{N_c}{(4\pi)^2} N_F \frac{\rho_0'}{144}, \\
H_4 &=& -\frac{N_c}{(4\pi)^2} N_F \frac{\rho_0'}{90},  \\ H_5
&=& -\frac{N_c}{(4\pi)^2} N_F \frac{7\rho_0'}{720}.
\end{eqnarray} 
Note that there is a finite strong renormalization
to Newton's gravitational constant $G$, since the classical Einstein's
Lagrangian is $ {\cal L} = - R / (16 \pi G)$. This correction,
proportional to the ratio of the hadronic to the Planck scale $f^2 G
\pi / 3 $, is numerically tiny. 

\subsection{Energy-Momentum tensor} 
\label{sec:ener-mom}

Using the action of Eq.~(\ref{eq:eff_ac}) one can compute the energy
momentum tensor as a functional derivative of the action with respect
to an external space-time-dependent metric, $ g_{\mu\nu} (x)$, around
the flat space-time metric $ \eta_{\mu \nu} $ (we take the 
signature $(+ - - -)$ ),
\begin{eqnarray}
\frac12 \theta^{\mu \nu} (x) &=& \frac{\delta \Gamma}{\delta g_{\mu
\nu}(x)} \Big|_{g_{\mu \nu} = \eta_{\mu\nu} } \\ &=& -{\rm
i} \frac{N_c}2 \int_C d \omega \rho(\omega) \langle x | \left\{ O^{\mu
\nu} \, , \, ( {\rm i} {\bf D} )^{-1} \right\} | x \rangle, \nonumber 
\label{eq:theta_quark}
\end{eqnarray} 
where 
\begin{eqnarray}
O^{\mu \nu} = \frac{\rm i}2 \left( \gamma^\mu \partial^\nu +
\gamma^\nu \partial^\mu \right) -  g^{\mu\nu} \left({\rm i}
\slashchar{\partial}- \omega \right) . 
\end{eqnarray} 
In the flat space-time limit, $ g^{\mu \nu} = \eta^{\mu \nu} $, the
chiral Lagrangean contains only metric contributions and takes the
form given in Ref.~\cite{Gasser:1983yg,Gasser:1984gg},
\begin{eqnarray}
{\cal L} = {\cal L}^{(0)}+{\cal L}^{(2)}+ {\cal L}^{(4)}+ \dots 
\label{eq:chil_flat} 
\end{eqnarray} 
where 
\begin{eqnarray} 
 {\cal L}^{(2)} &=& {\cal L}^{(2,g)} \Big|_{g_{\mu \nu} =
 \eta_{\mu\nu} }, \nonumber \\ {\cal L}^{(4)} &=& {\cal L}^{(4,g)}
 \Big|_{g_{\mu \nu} = \eta_{\mu\nu} }.
\end{eqnarray} 
If we do a derivative expansion (see Appendix \ref{sec:app} for
details) the effective chiral energy-momentum tensor up to and
including fourth order corrections in the chiral counting
reads~\cite{Donoghue:qv}
\begin{eqnarray}
\theta_{\mu \nu} = \theta_{\mu \nu}^{(0)}+\theta_{\mu \nu}^{(2)} +
\theta_{\mu \nu}^{(4)} + \dots 
\end{eqnarray} 
where  
\begin{eqnarray}
\theta_{\mu \nu}^{(0)} &=& -g_{\mu \nu} {\cal L}^{(0)} , \\
\theta_{\mu \nu}^{(2)} &=& \frac{f^2}4 \langle D_\mu U^\dagger D_\nu U
\rangle - g_{\mu\nu}{\cal L}^{(2)}, \label{eq:en-mom}  \\ 
\theta_{\mu \nu}^{(4)} &=& - g_{\mu\nu}{\cal L}^{(4)} + 2 L_4 \langle
D_\mu U^\dagger D_\nu U \rangle \langle \chi^\dagger U + U^\dagger \chi
\rangle \nonumber \\ &+& L_5 \langle D_\mu U^\dagger D_\nu U + D_\nu
U^\dagger D_\mu U \rangle \langle \chi^\dagger U + U^\dagger \chi
\rangle \nonumber \\ &-& 2 L_{11}\left( g_{\mu \nu} \partial^2 -
\partial_\mu \partial_\nu \right) \langle D_\alpha U^\dagger D^\alpha
U^\dagger \rangle \nonumber \\ &-& 2 L_{13} \left(g_{\mu \nu}
\partial^2 - \partial_\mu \partial_\nu \right) \langle \chi^\dagger U
+ U^\dagger \chi \rangle \nonumber \\ &-& L_{12} \left( g_{\mu\alpha}
g_{\nu \alpha} \partial^2 + g_{\mu\nu} \partial_\alpha \partial_\beta
- g_{\mu \alpha} \partial_\nu \partial_\beta - g_{\nu \alpha}
\partial_\mu \partial_\beta \right) \nonumber \\ && \times \langle
D^\alpha U^\dagger D^\beta U \rangle . \nonumber \\
\end{eqnarray} 
Note that the coefficients $L_1-L_{10} $ appear in ${\cal L}^{(4)}$
given by Eq.~(\ref{eq:chl4}).  The terms containing $L_{11}-L_{13}$
cannot be obtained by computing the energy momentum tensor from the
chiral effective Lagrangean in flat-space time (\ref{eq:chil_flat})
and form this viewpoint are genuine quark contributions to
$\theta^{\mu \nu}$ in this model.  Actually, the difference between
computing the energy-momentum tensor from an action at the quark, {\em i.e.}
starting from Eq.~(\ref{eq:theta_quark}), or  at the meson level,
{\em i.e.} starting from Eq.~(\ref{eq:chil_flat}) is
\begin{eqnarray} 
\frac{\delta \Gamma}{\delta g_{\mu \nu}(x)} \Big|_{g_{\mu \nu}}-
\frac{\delta S^g}{\delta g_{\mu \nu}(x)} \Big|_{g_{\mu \nu}} =
\frac{\delta S^R}{\delta g_{\mu \nu}(x)} \Big|_{g_{\mu \nu}},
\end{eqnarray} 
with $S^g$ and $S^R$ denoting the metric and curvature contributions to the
action, is precisely related to the curvature terms corresponding
to the couplings $L_{11}, L_{12}$, and $L_{13}$.

\section{Results for the Meson Dominance Model} 
\label{sec:md}

The Meson Dominance Model (MDM), developped in
Ref.~\cite{RuizArriola:2003bs}, offers a particularly simple 
realization of the SQM and provides an explicit form for the 
spectral function. The quark propagator becomes
\begin{eqnarray}
S({p}) = \int_C d \omega \frac{ \rho_V (\omega) \slashchar{p} +
\rho_S(\omega)\omega }{ p^2 - \omega^2 } = \frac{Z(p^2) }
{\slashchar{p} -M(p^2)},
\label{eq:spec1} 
\end{eqnarray} 
where 
\begin{eqnarray}
\rho_V (\omega) &=& \frac{1}{2\pi i} \frac{1}{\omega}
\frac{1}{(1-4\omega^2/M_V^2)^{5/2}}, \\\rho_S (\omega) &=& 
\frac{1}{2\pi i} \frac{12 \rho^\prime_3} {M_S^4
(1-4\omega^2/M_S^2)^{5/2}} . \label{rhos}
\end{eqnarray} 
The vector spectral function, $ \rho_V (\w) $, is determined by
imposing vector meson dominance of the pion 
electromagnetic form factor, from
which the identity  
\begin{eqnarray}
f^2 = \frac{N_c M_V^2}{24\pi^2}  
\label{eq:fpi_mrho}
\end{eqnarray} 
is deduced. This relation is subject to chiral corrections. It is
remarkable that such a simple relation produces a mass of $M_V = 826
{\rm MeV} $ for $f_\pi= 93 {\rm MeV} $ which agrees with the value
recently obtained in Ref.~\cite{Pelaez:2003dy}. With this value of $f$
one gets a vacuum energy of $B=-3N_F f^4 / N_c \sim (202-217 {\rm
MeV})^4 $ for $N_F=3$. In contrast to $\rho_V(\w)$, the expression
for the scalar spectral function, $\rho_S ( \w) $, is an educated
guess which fulfills the odd spectral conditions $ \rho_1 = \rho_3 =
\dots =0$ and reproduces the value of the $\rho^\prime_3 $ log-moment.
The preferred value for the vector mass is
\begin{equation}
M_V=770 {\rm MeV},
\end{equation} which corresponds to the $\rho$-meson mass and 
which is used in the subsequent numerical analysis.

The integration contour $C$ used in the MDM encircles the branch cuts,
{\em i.e.}, starts at $-\infty+i0$, goes around the branch point at
$-M_V/2$, and returns to $-\infty -i0$, with the other section
obtained by a reflexion with respect to the
origin~\cite{RuizArriola:2003bs}.  These two sections are connected
with semicircles at infinity. The mass function becomes
\begin{eqnarray} 
\frac{M(p^2)}{M(0)} &=& \frac{ 10 p^2}{M_V^2}\frac{
\left(\frac{M_S^2}{M_S^2 -4 p^2} \right)^{5/2}}{
\left(\frac{M_V^2}{M_V^2 -4 p^2} \right)^{5/2} -1} ,
\end{eqnarray} 
where the constituent quark mass is\footnote{In Ref.~\cite{RuizArriola:2003wi} there were 
typographical errors in Eq.~(10.6) and (10.9), which should carry an extra factor of 2
on the RHS.}
\begin{eqnarray}
M_Q \equiv M(0) = - \frac{48 M_V^2 \pi^2 \langle \bar q q \rangle }{5 M_S^4 N_c}.
\end{eqnarray} 
When $M(p^2)=p^2$ then $Z(p^2)=0 $, such that the quark propagator has
no poles in the complex $p^2$-plane. Instead, it has a cut starting at
the branch-point $p^2 = M_V^2 / 4$. The exponents reproduce accurately
the $ 1/(-p^2)^{3/2} $ behaviour in the deep-Euclidean domain. This
behavior was seen in the recent QCD lattice simulation in the Landau
gauge, linearly extrapolated to the chiral
limit~\cite{Bowman:2002bm}. A fit to the data yields
\cite{RuizArriola:2003wi}
\begin{eqnarray}
M_Q &=& 303 \pm 24~{\rm MeV}, \nonumber \\
M_S &=& 970 \pm 21~{\rm MeV}, \label{msmq}
\end{eqnarray} 
with the optimum value of $\chi^2$ per degree of freedom equal to
0.72, yielding and impressive agreement of $M(p^2)$ up to $ p^2 = -16
{\rm GeV}^2$.  Although $Z(p^2)$ is not nearly as good ({\em cf.}
Ref.~\cite{RuizArriola:2003wi}), leaving room for improvement, we
think it worthwhile to pursue the pattern of chiral symmetry breaking
which arises in this particular realization of the SQM. Incidentally,
let us note that if the results of Sect.~\ref{sec:moments} are used we
get
\begin{eqnarray}
f^2 = \frac{N_c}{4 \pi^2} \int d p_E^2 \frac1{\left(1+ \frac{4
p_E^2}{M_V^2}\right)^{5/2} },
\end{eqnarray}
which reproduces Eq.~(\ref{eq:fpi_mrho}) and shows the consistency of the
approach.  For the Meson Dominance model we get
\begin{eqnarray}
{\rho^\prime_1}^{\rm MD} &=& \frac{ 8 \pi^2 \langle \bar q q \rangle
}{ N_c M_S^2} = - \frac{5 M_Q M_S^2 }{6 M_V^2}, \nonumber \\
{\rho^\prime_2}^{\rm MD} &=& - \frac{4 \pi^2 f^2 }{N_c} = - \frac{M_V^2}{6} ,\\ 
{\rho^\prime_3}^{\rm MD} &=& - \frac{ 4 \pi^2 \langle \bar q q \rangle
}{N_c}= \frac{5 M_Q M_S^4}{12 M_V^2} \nonumber .
\label{eq:md}
\end{eqnarray} 
Using these values we get 
\begin{eqnarray}
L_5 &=& \frac{N_c}{96 \pi^2} \frac{M_V^2} {M_S^2} ,\\ L_7 &=&
\frac{N_c}{32 \pi^2 N_f} \left( \frac1{12} -\frac{M_V^2} {6 M_S^2}
\right), \\ L_8 &=& \frac{N_c}{16 \pi^2}\left( - \frac{M_V^{10}}{150
M_Q^2 M_S^8} + \frac{M_V^2}{12 M_S^2} - \frac1{24} \right) .
\end{eqnarray} 
In the SU(3) case we display our results in Table~\ref{tab:table2}. We
note that the predictions for $L_{1,2,3,4,6,9,10} $ are common to the
scheme of Ref.~\cite{Espriu:1989ff}. The values of $ L_{5,7,8}$ are
specific both to the SQM and the MDM realization.  

\begin{table}
\caption{\label{tab:table2} The dimensionless low energy constants
(multiplied by $10^3$) compared with some reference values and other
models. The errors for SQM in the MDM realization reflect the errors in 
$M_S$ and $M_Q$ of Eq.~(\ref{msmq}).}
\begin{ruledtabular}
\begin{tabular}{ccccccc}
$\times 10^3$ & SQM & ChPT\footnotemark[1] & Large
$N_c$ \footnotemark[2] & NJL\footnotemark[3] & Dual \\
 & (MDM) & & & & Large $N_c$ \\
\hline
$L_1$ & 0.79 & 0.53$\pm$0.25 & 0.9 & 0.96 &  0.79  \\
$L_2$ & 1.58 & 0.71$\pm$0.27 & 1.8 & 1.95 &  1.58  \\
$L_3$ & -3.17 & -2.72$\pm$1.12 & -4.3 & -5.21 & -3.17  \\
$L_4$ & 0 & 0  & 0 & 0 & 0 \\
$L_5$ & 2.0$\pm$0.1 & 0.91$\pm$0.15 & 2.1 & 1.5 &  3.17 \\
$L_6$ & 0 & 0 & 0 & 0 & 0  \\
$L_7$ & -0.07$\pm$0.01 \footnotemark[4] & -0.32$\pm$0.15 & -0.3 &  &    \\
$L_8$ & 0.08$\pm$0.04  & 0.62$\pm$0.20 & 0.8 & 0.8 & 1.18  \\
$L_9$ & 6.33 & 5.93$\pm$0.43 & 7.1 & 6.7 &  6.33  \\
$L_{10}$& -3.17 & -4.40$\pm$0.70\footnotemark[5] & -5.4 & -5.5 &  -4.75  \\
$L_{11}$& 1.58 & 1.85$\pm$0.90\footnotemark[6] & 1.6 &  &   \\
$L_{12}$& -3.17 & -2.7 \footnotemark[6] &  -2.7 &  &   \\
$L_{13}$& 0.33$\pm$ 0.01 & 1.7 $\pm$0.80\footnotemark[6] & 1.1 &  &    \\
\end{tabular}
\end{ruledtabular}
\footnotetext[1]{The two-loop calculation of Ref.~\cite{Amoros:2001cp}}
\footnotetext[2]{Ref.~\cite{Ecker:1988te}}
\footnotetext[3]{Ref.~\cite{RuizArriola:gc}}
\footnotetext[4]{See footnote \ref{foot}}
\footnotetext[5]{Ref.~\cite{Bijnens:2002hp,Bijnens:1996wm}}
\footnotetext[6]{Ref.~\cite{Donoghue:qv}}
\end{table}

In the SU(2) case we have, with the help of the relations given in
Ref.~\cite{Gasser:1984gg}, to pass form SU(3) to
SU(2)~\cite{Gasser:1983yg}. In the absence of meson loop
corrections~\footnote{The relations are $ \bar l_1
= 192 \pi^2 ( 2 L_1 + L_3) $, $\bar l_2 = 192 \pi^2 L_2 $, $\bar l_3 =
256 \pi^2 ( 2 L_4 + L_5 - 4 L_6 - 2 L_8 ) $, $\bar l_4 = 64 \pi^2 ( 2
L_4 + L_5) $, $\bar l_5 = - 192 \pi^2 L_{10}$, $ \bar l_6 = 192 \pi^2
L_9 $, $\bar l_{11} = 192 \pi^2 L_{11}$ , $\bar l_{13} = 256 \pi^2 l_{13}$. The constant
$l_{12}$ is not renormalized by the pion loop.}
\begin{eqnarray} 
\bar l_1 &=& - \bar l_2 = - \frac12 \bar l_5 = - \frac14 \bar l_6 =
-N_c, \\ 
\bar l_3 &=& \frac{4 N_c }3 + \frac{16 N_c M_V^{10}}{75 M_Q^2 M_S^8}, \\
\bar l_4 &=& \frac{ 2 M_V^2 N_c }{3 M_S^2}. 
\end{eqnarray} 
The vector and scalar pion radii are  given by~\cite{Gasser:1983yg} 
\begin{eqnarray}
\langle r^2 \rangle_V &=& \frac1{16\pi^2 f^2 } \bar l_6 =\frac{6}{M_V^2}, \nonumber \\
\langle r^2 \rangle_S &=& \frac{3}{8 \pi^2 f^2} \bar l_4 =\frac{6}{M_S^2},
\label{eq:radii} 
\end{eqnarray} 
respectively. While the vector pion mean squared radius reproduces the
built-in vector meson dominance of the pion e.m. form factor, the
scalar radius shows that the scalar mass obtained by a fit to the
lattice quark mass function does correspond to the mass of a scalar
meson dominating the scalar form factor, 
$\langle r^2 \rangle^{1/2}_S = 0.50 \pm 0.01 {\rm fm} $.

The scalar (spin-0) and tensor (spin-2) component of the gravitational
form factors, $\theta_0 $ and $\theta_2$~\cite{Donoghue:qv},
respectively, produce the same mean squared radii,
\begin{eqnarray}
\langle r^2 \rangle_{G,0} &=&  \langle r^2 \rangle_{G,2} =
\frac{N_c}{48\pi^2 f^2 }, 
\end{eqnarray}
regardless of the particular realization of the spectral model.  If we
saturate the form factors with scalar and tensor mesons, $f_0$ and
$f_2$, we 
 get for their masses
\begin{eqnarray}
M_{f_0}= M_{f_2} = 4 \pi f_\pi \sqrt{3/N_c} = 1105-1168 {\rm MeV}.  
\end{eqnarray} 
depending whether we take $f=88$ or $93 {\rm MeV}$, respectively. The
experimental value for the lowest tensor meson is $M_{f_2}^{\rm exp.}
= 1270 {\rm MeV} $. As discussed in Ref.~\cite{Donoghue:qv}, the
$\theta_0 $ (corresponding to the trace of the energy-momentum tensor)
form factor couples to scalars, whereas the $ \theta_2 $ 
(corresponding to the traceless combination of $\theta_{\mu\nu} $)
form factor couples to tensor (spin-2) mesons. 

One message is clear from the present model: the scalar meson of mass
$M_{f_0}$ which dominates the energy-momentum tensor does not
necessarily coincide with the scalar meson of mass $M_S$, which
dominates the scalar form factor. Actually we have $M_{f_0} = \sqrt{2}
M_V $, whereas $M_S $ is a free quantity. This is natural in the
spectral approach, where in the chiral limit the scalar form factor
$F_S$ involves the odd spectral moments, whereas $\theta_0 $ involves
the even spectral moments. In particular, the corresponding mean
squared radii are proportional to $\rho'_1$ and to $\rho_0$,
respectively. Finally, we note the numerical value of $\bar l_3=4.65 $
obtained in MDM amounts to a shift of the pion mass by less than $1\%
$ and an increase of $ f_\pi $ yielding $ 89 {\rm MeV} $ as compared
to $ f = 87 {\rm MeV} $.

\section{The large-$N_c$ limit and duality} 
\label{sec:dual}

Given the fact that our result corresponds to a one-quark-loop
approximation, we cannot expect our model to be better than the leading
large-$N_c$ contribution to the low-energy parameters, which is made
of infinitely many resonance exchanges~\cite{Pich:2002xy}. On the
other hand, the evaluation of these large-$N_c$ contributions requires
additional, not necessarily unreasonable, assumptions such as
the convergence of an infinite set of states, and moreover, an estimate
of the contributions of higher resonances. In practice, one works in
the Single Resonance Approximation (SRA) yielding a reduction of
parameters~\cite{Pich:2002xy,Donoghue:qv},
\begin{eqnarray} 
2 L_1^{\rm SRA} &=& L_2^{\rm SRA} = \frac14 L_9^{\rm SRA} = - \frac13
L_{10}^{\rm SRA} = \frac{f^2}{8 M_V^2},  \\ L_5^{\rm SRA} &=&
\frac{8}{3} L_8^{\rm SRA} = \frac{f^2}{4 M_S^2}, \\ L_3^{\rm SRA} &=& -
3 L_2^{\rm SRA} + \frac12 L_5^{\rm SRA},  \\ 
2 L_{13}^{\rm SRA} &=& 3 L_{11}^{\rm SRA} + L_{12}^{\rm SRA} =
\frac{f^2}{4 M_{f_0}^2}, \\ 
L_{12}^{\rm SRA} &=& -\frac{f^2}{2 M_{f_2}^2}, 
\label{eq:SRA}
\end{eqnarray} 
where $f$, $M_V$ and $M_S$ should stand for the leading large-$N_c$
contributions to those quantities. To obtain the formulas for $L_1$
till $L_{10} $, the pseudoscalar and axial meson contributions have
been fine tuned to satisfy the VV-AA and SS-PP two point correlation
functions high-energy-behavior chiral sum rules plus some well
converging high energy properties of hadronic form factors. (In
particular, $M_P /M_S = M_A / M_V = \sqrt{2}$, where $M_P$ is the mass
of the excited pion. Obviously, more short-distance constraints
require more resonances.  The values of $L_{11,12,13} $ are obtained
from the single scalar and tensor resonance
exchange~\cite{Donoghue:qv}. On the one hand, a tensor meson is needed
in order to provide a non-vanishing $L_{12}$ as a minimal hadronic
ansatz, on the other hand tensor mesons do contribute also other
LEC's~\cite{Toublan:1995bk}, which is not taken into account in
Eq.~(\ref{eq:SRA}). Thus, to simplify the discussion, in what follows
we restrict ourselves to the non-gravitational couplings $L_1$ till
$L_{10}$. In practice, the phenomenological success is achieved by
using the physical values of the parameters.  Note that although there
is predictive power, it is done in terms of two dimensionless ratios,
$ f/M_V $ and $f /M_S $. Obviously, in the chiral limit we expect both
$ M_V $ and $M_S $ to scale with $f_\pi$. Therefore, in order to
preserve the large-$N_c$ counting rules one should have $ M_V = c_V
f_\pi /\sqrt{N_c} $ and $ M_S = c_S f_\pi /\sqrt{N_c} $ with $ c_V $
and $c_S $ denoting some $N_c$-independent coefficients. Remarkably,
in the SQM the low energy parameters depend on two dimensionless
ratios, $ \rho^\prime_1 / B_0 $ and $ \rho'_2 / B_0^2$. It is
therefore tempting to determine the spectral log-moments from large
$N_c$ arguments, in a model-independent way. Actually in the Single
Resonance Approximation (SRA) we note that the ratios $ L_1:L_2:L_9 $
of the SQM agree with those of the SRA. The values of $L_5$ and $ L_9$
can then be used to determine $\rho_1^\prime $, and $\rho_2^\prime $,
respectively, yielding
\begin{eqnarray}
{\rho^\prime_1}^{\rm SRA} &=&  \frac{ 8 \pi^2 \langle \bar q q \rangle }{ N_c
M_S^2}, \\
{\rho^\prime_2}^{\rm SRA} &=& - \frac{4 \pi^2 f^2 }{N_c} = -
\frac{M_V^2}{6} ,
\label{eq:rho_sra} 
\end{eqnarray}
in agreement with Eqs.~(\ref{eq:md}) and Eq.~(\ref{eq:fpi_mrho}). This
is not surprising since the physics of the meson dominance version of
the SQM and the SRA approximation is alike. The only difference is
that one cannot deduce from Eqs.~(\ref{eq:rho_sra}) the value of the
constituent quark mass $M_Q=M(0)$, which is given by the ratio of two
negative moments $ M_Q= \rho_{-1} / \rho_{-2} $,
Eq.~(\ref{eq:q_mass}). To determine $M_Q$ would require computing terms
of ${\cal O} (p^6 ) $ in the chiral Lagrangian and comparing to the
SRA at large $N_c$.

One can see that it is not possible to match $L_8$ nor $L_{10}$.  The
disagreement with the large-$N_c$ values of $L_{8}$ and $L_{10}$ has
to do with the fact that the SS-PP sum rule and VV-AA second
Weinberg sum rule are violated in the present as well as other quark
model calculations \cite{Peris:1998nj,Bijnens:2003rc} (except for the non-local
models, see \cite{WBcoim,Dorokhov:2003kf}).  This calls for a modification of
our model. The disagreement has to do with the absence of
axial-meson exchange in $L_{10}$ (1/4 of the total contribution) and
pseudoscalar meson exchange in $L_8$ (1/4 of the total
contribution). On the other hand, for the value of $f$ obtained from
Eq.~(\ref{eq:fpi_mrho}) the constants $L_1$, $L_2$, $L_4$, $L_5$, $L_6$, $L_9$
reproduce the large-$N_c$ constraints obtained in
Ref.~\cite{Ecker:1988te}. This agreement is confirmed in
Table~\ref{tab:table2} if one corrects for the 
factor $ 24\pi^2 f_\pi^2 / N_c M_V^2 = 1.15$. One could force $L_3$ to
agree to the large-$N_c $ estimate by taking $M_V=M_S$. This agrees
with the observation of the Chiral Unitary approach of
Ref.~\cite{Pelaez:2003dy}; in the large-$N_c $ limit the scalar and
vector mesons become degenerate~\footnote{For $N_c=3,10,20,40$
Ref.~\cite{Pelaez:2003dy} obtains $M_S /M_V = 0.58,0.84,0.96, 0.98 $
respectively with $M_S$ and $M_V$ the real part of the poles in the
second Riemann sheet. We thank J.R. Pel\'aez for providing these
numbers.}. Thus, the marriage of large-$N_c $ in SRA approximation
with our chiral quark model calculation produces degenerate scalar and
vector mesons. Degenerated scalar and vector mesons were suggested
very early \cite{Gilman:1967qs} in the context of superconvergent sum
rules and have been interpreted more recently on the basis of mended
symmetries~\cite{Weinberg:xn}. Experimental claims have been
raised~\cite{Svec:1996xp,Svec:2002bz,Svec:2002rw} and
contested~\cite{Kaminski:2000bd}.  Direct experimental tests have also
been suggested~\cite{Ahmedov:2003sb}.

It is clear that whatever sensible modification of the SQM is
considered, it will only affect $L_8$ and $L_{10} $, keeping the
remaining $L's$. We leave the explicit construction of such a modified
model for a separate study. Regardless on the particular way to
achieve this, we may anticipate already on the consequences for the
large $N_c$ in the single resonance approximation of taking $M_S=M_V=
2 \pi f \sqrt{6/N_c} $, yielding the following duality relations,
\begin{eqnarray} 
2 L_1 &=& L_2 = - \frac12 L_3 =\frac12 L_5 = \frac23 L_8= \frac14 L_9 =
- \frac13 L_{10} \nonumber \\ &=& \frac{N_c}{192\pi^2} .
\label{eq:L_dual} 
\end{eqnarray} 
This also implies the set of mass dual relations,
\begin{eqnarray}
M_A = M_P = \sqrt{2} M_V = \sqrt{2} M_S = 4 \pi \sqrt{3/N_c} f_\pi \, .   
\end{eqnarray} 
The new relation $M_A = M_P $ agrees with the experimental number
within the expected $30 \%$ of the large-$N_c$ limit.  Using
Eqs.~(\ref{eq:radii}) we obtain
\begin{eqnarray}
\langle r^2 \rangle^{1/2}_S = \langle r^2 \rangle^{1/2}_V = 2
\sqrt{N_c} / f_\pi .
\end{eqnarray}    
These relations are subject to higher $1/N_c$ and $m_\pi $
corrections. We may account for the latter by allowing $ f_\pi $ to
vary between the physical value and the value in the chiral
limit. This yields, $ \langle r^2 \rangle^{1/2}_S = \langle r^2
\rangle^{1/2}_V = 0.58-0.64~{\rm fm} $. The value of the scalar radius
is compatible with the one obtained in ChPT to two
loop~~\cite{Colangelo:2001df}, $ 0.78~{\rm fm} $. Going to the SU(2)
case, in the dual large-$N_c$ model we get
\begin{eqnarray}
- \bar l_1 =  \bar l_2 = \frac32 \bar l_3 = \frac32 \bar l_4  =
  \frac13 \bar l_5 = \frac14 \bar l_6 = N_c ,
\end{eqnarray} 
whereas the recently extracted values obtained at the two loop level
from the analysis of $\pi\pi$ scattering~\cite{Colangelo:2001df} and
vector and scalar form factors~\cite{Bijnens:1998fm} at the two loop
level are 
\begin{eqnarray} 
\bar l_1 = -0.4 \pm 0.6 \, , \, 
\bar l_2 &=& 6.0 \pm 1.3  \, , \, 
\bar l_3 = 2.9 \pm 2.4,  \nonumber \\ 
\bar l_4 = 4.4 \pm 0.2  \, , \, 
\bar l_5 &=& 13.0 \pm 1.0 \, , \, 
\bar l_6 = 16.0 \pm 1.0  .  \nonumber \\
\end{eqnarray} 
The $\bar l$ coefficients are in a sense more suitable for
comparison with ChPT since the chiral loop generates a constant shift
in all of them by the same amount, $c=\log ( \mu^2 / m^2)$. Thus, it
makes sense to compare the differences where chiral logs are
canceled. We find
\begin{eqnarray} 
\bar l_2 -\bar l_1 &=& 2 N_c \quad ( {\rm Exp.}\, 6.4 \pm 1.4 ),
\nonumber \\ \bar l_3 -\bar l_1 &=& \frac{5N_c}3 \quad ( {\rm Exp.}\,
3.3 \pm 2.4 ), \nonumber \\ \bar l_4 -\bar l_1 &=& \frac{5N_c}3 \quad
( {\rm Exp.}\, 4.8 \pm 0.4 ), \\ \bar l_5 -\bar l_1 &=& 4 N_c \quad (
{\rm Exp.}\, 13.4 \pm 1.1 ), \nonumber \\ \bar l_6 -\bar l_1 &=& 5 N_c
\quad ( {\rm Exp.}\, 16.4 \pm1.1 ) \nonumber ,
\end{eqnarray} 
where the errors have been added in quadrature. As we can see, the
agreement is excellent, within the uncertainties, and suggests
accuracy 
 of the order of $ 1/N_c^2$ rather than the standard a
priori $ 1/N_c$ error estimate. The constant pion loop shift can be
accommodated with a scale $ \mu = 513\pm 200 {\rm MeV}$, comparable to
the $\rho$ meson mass. Taking Eqs.~(\ref{eq:SRA}), corresponding to
the SRA with the physical values $ f=93 {\rm MeV} $ , $M_S = 1000 {\rm
MeV} $, and $ M_V = 770 {\rm MeV} $, as done in
Ref.~\cite{Pich:2002xy}, yields $ \bar l_2 - \bar l_1 = 8.3$, $\bar
l_3 - \bar l_1 = 6.2$, $\bar l_4 - \bar l_1 = 6.2$, $\bar l_5 - \bar
l_1 = 15.2$, $\bar l_6 - \bar l_1= 18.7$. More reasonable values are
obtained by taking $M_S = 600 {\rm MeV} $, but then the SRA relation
$M_P = \sqrt{2} M_S $ predicts a too low value of the excited pion
state. The present discussion favours phenomenologically the dual
relations (\ref{eq:L_dual}) as compared to the SRA relations
(\ref{eq:SRA}) with physical parameters.

\section{Conclusions} 

In the present work we have studied the chiral expansion of the
recently proposed Spectral Quark Model in the presence of electroweak
and gravitational external sources. The model is based on a Lehman
representation for the quark propagator with an unconventional spectral
function, which is genuinely a complex function with cuts in terms of
the spectral mass. We have written down the effective action which reproduces
the Ward-Takahashi identities presented in the previous work. Thanks
to an infinite set of spectral conditions demanded from the power like
factorization property of form factors at high energies, we have been
able to show that the corresponding chiral anomalous contribution to
the action is properly normalized without removing the
regularization. Moreover, the non-anomalous contribution to the action
can be written in the long wavelength limit in terms of 13 low energy
constants. The numerical values are in reasonable agreement with the
phenomenological expectations, although some discrepancies do occur
for $L_8$ and $L_{10}$. In some cases they can be naturally explained
as failures in reproducing some chiral short distance constraints
which suggest that the model needs to be improved. On the other hand,
if one tries to match the remaining non-gravitational LEC's to large
$N_c$ predictions in the single resonance approximation, a further
reduction of parameters takes place. In particular, one finds 
the best agreement for
degenerate scalar and vector mesons. 

We have estimated for the first time in the framework of chiral
quark models the gravitational LEC's $L_{11}$,
$L_{12}$ and $L_{13}$, describing the coupling to external gravitational
sources. These LEC's depend on curvature properties of the curved
space-time metric. This calculation allows a determination of some
matrix elements of the energy momentum tensor. Our analysis suggests
that the scalar meson coupling to the quark condensate $ m_0 \bar q q
$ and the scalar meson coupling to the trace of the energy momentum
tensor $\theta^\mu_\mu$ do not necessarily coincide. Clearly, these
two operators behave differently under chiral symmetry, since $m_0
\bar q q $ vanishes in the chiral limit whereas $\theta^\mu_\mu$ does
not. This point is in itself rather intriguing and deserves further
investigation. We note here that this fact materializes in our model
because these two scalar mesons depend on odd and even spectral
moments respectively.  On the other hand, we obtain $M_{f_0}=M_{f_2}=
\sqrt{2} M_V = \sqrt{2} M_S = 4\pi \sqrt{3/N_c} f_\pi $, a very
reasonable result if we take into account the large $N_c $ nature of
the one quark loop approximation. Further quark-meson duality
relations have been discussed, allowing a rather successfull
determination of the best known LEC's, consistent up to the
experimental errors with the best known values up to two loop
accuracy.

\begin{acknowledgments}

We thank Valery Lubovitskiy for reminding us of the work of Efimov and
Ivanov. One of us (E.R.A.) thanks J. Prades, S. Peris and
J. R. Pel\'aez for useful discussions and correspondence. This work is
supported in part by funds provided by the Spanish DGI with grant
no. BMF2002-03218, Junta de Andaluc\'{\i}a grant no. FM-225 and
EURIDICE grant number HPRN-CT-2003-00311. Partial support from the
Spanish Ministerio de Asuntos Exteriores and the Polish State
Committee for Scientific Research, grant number 07/2001-2002 is also
gratefully acknowledged.
\end{acknowledgments}

\appendix

\section{Derivative Expansion and Useful Identities} 
\label{sec:app}

\subsection{Reduction to a vector like theory and transformation properties} 

The Dirac operator can be rewritten as
\begin{eqnarray} 
\D = \D_R P_R + \D_L P_L,
\end{eqnarray} 
with the projection operators on parity
\begin{eqnarray} 
P_R = {1\over2} ( 1+ \gamma_5 ) , \qquad
P_L = {1\over2} ( 1- \gamma_5 ),
\end{eqnarray} 
such that for a Dirac spinor one has 
\begin{eqnarray} 
\Psi_R = P_R \Psi , \qquad
\Psi_L = P_L \Psi .
\end{eqnarray} 
The right and left Dirac operators are given by
\begin{eqnarray} 
i \D_R &=& i \slashchar\partial + {\slashchar A}_R - {\cal M} ,
\nonumber \\ 
i \D_L &=& i \slashchar\partial + {\slashchar A}_L - {\cal M}^\dagger , 
\end{eqnarray} 
with
\begin{eqnarray} 
{\cal M} = s + i p + \w U   &,&  \qquad  
{\cal M}^\dagger = s - i p + \w U^\dagger,   \\
A^\mu_R = v^\mu + a^\mu  &,& \qquad  A^\mu_L = v^\mu - a^\mu .
\end{eqnarray} 
the quark mass matrix is included in the scalar field $s$. Under
left-right unitary transformations, $\Omega_L$ and $\Omega_R$, one has
the following properties, 
\begin{eqnarray}
\Psi_R &\to& \Omega_R \Psi_R, \qquad \Psi_L \to \Omega_L \Psi_L ,\\
U &\to& \Omega_L U \Omega_R^\dagger, \qquad U^\dagger \to
\Omega_R U^\dagger \Omega_L^\dagger ,\\
A^\mu_R &\to& \Omega_R A^\mu_R
\Omega_R^\dagger + i \Omega_R \partial^\mu  \Omega_R^\dagger ,\\ 
A^\mu_L &\to& \Omega_L A^\mu_L \Omega_L^\dagger + i \Omega_L
\partial^\mu \Omega_L^\dagger .
\end{eqnarray} 
The chiral covariant derivatives and field strength tensors 
\begin{eqnarray} 
D_\mu \Psi_R &=& \partial_\mu \Psi_R - i A_\mu^R \Psi_R , \nonumber \\ 
D_\mu U &=& D_\mu^L U-U D_\mu^R =
\partial_\mu U-i A_\mu^L U + i U  A_\mu^R , \nonumber \\ 
 F_{\mu\nu}^r &=& i[ D_\mu^r, D_\nu^r] = 
\partial_\mu A_\nu^r -\partial_\nu A_\mu^r
-i [ A_\mu^r , A_\nu^r ], \nonumber \\
&& \;\;\; r=R,L . 
\end{eqnarray} 
behave as follows under local chiral transformations: 
\begin{eqnarray} 
D_\mu \Psi_R &\to& \Omega_R D_\mu \Psi_R, \\ D_\mu \Psi_L &\to&
\Omega_L D_\mu \Psi_R, \\ D_\mu U &\to& \Omega_L D_\mu U
\Omega_R^\dagger, \\ D_\mu U^\dagger &\to& \Omega_R D_\mu U^\dagger
\Omega_L^\dagger.
\end{eqnarray}

\subsection{Coupling the Spectral Quark Model to Gravity} 

The coupling of fermions to gravity is well known (see,
{\em e.g.}, Ref.~\cite{Birrell:ix}) but not in the context of chiral quark
models. We review it here for completeness and to fix our notation. 
We use the tetrad formalism of curved space-time (for
conventions see, {\em e.g.}, Ref.~\cite{We-grav}). Given the metric tensor we
get a local basis of orthogonal vectors (tetrads or vierbein),
\begin{eqnarray}
g^{\mu \nu} (x) = e^\mu_A  (x) e^\nu_B (x) \eta^{AB}  ,
\end{eqnarray} 
with $\eta^{AB} = {\rm diag} ( 1, -1 , -1 , -1 ) $ for a flat Minkowski
metric. These vectors fulfill the orthogonality relations,
\begin{eqnarray}
\delta^\mu_\nu &=& \eta^{AB} e^\mu_A e_{\nu B} = e^\mu_A e^A_\nu , \nonumber \\
\delta^A_B &=& g^{\mu \nu} e_\mu^A e_{\nu B} = e_\mu^A e_B^\mu .
\end{eqnarray}  
Under the coordinate $ x^\mu \to x'^\mu (x) $ and frame $ x^A \to
\Lambda^A_B x^B $ transformations the transformation properties of the
tetrad are
\begin{eqnarray}
e_\mu^A \to \frac{\partial x'^\nu}{\partial x^\mu}  e_\nu^A , \nonumber \\
e_\mu^A \to \Lambda^A_B (x) e_\mu^B , 
\label{eq:tetrads}
\end{eqnarray} 
respectively. The tetrads map coordinate tensors into frame tensors
(which transform covariantly under local Lorentz transformations), for
instance
\begin{eqnarray}
T^{AB} = e^A_\mu e^B_\nu T^{\mu \nu} . 
\end{eqnarray} 
Frame tensors are invariant under coordinate transformations $ x^\mu
\to x'^\mu $. For a general tensor $ T_{\nu A}^\alpha $ greek indices
transform covariantly under coordinate transformations while latin
indices transform covariantly under frame transformations according to
Eq.~(\ref{eq:tetrads}) as follows,
\begin{eqnarray}
T_{\nu A}^\alpha \to \frac{\partial x'_\nu}{\partial x^\mu}
\frac{\partial x'^\alpha}{\partial x^\beta} \Lambda_A^B (x) T_{\mu
B}^\beta .
\end{eqnarray}  
The covariant derivative is defined as
\begin{eqnarray}
d_\mu T_{\nu A}^\alpha &=& \partial_\mu T_{\nu A}^\alpha - \Gamma_{\nu
\mu}^\lambda T_{\lambda A}^\alpha + \Gamma_{\mu \lambda}^\alpha
T_{\nu A}^\lambda+ \omega_{AB\mu} T_{\nu}^{\alpha B},  \nonumber \\ 
\end{eqnarray} 
where the Riemann connection is given by the Christoffel symbols, 
\begin{eqnarray}
\Gamma_{\lambda \mu}^\sigma = \frac12 g^{\nu \sigma} \left\{
\partial_\lambda g_{\mu\nu} + \partial_\mu g_{\lambda \nu} -
\partial_\nu g_{\mu \lambda} \right\} ,
\label{eq:christoffel}
\end{eqnarray} 
which are symmetric in the lower indices, 
$ \Gamma_{\lambda \mu}^\sigma = \Gamma_{\mu \lambda}^\sigma$ (we
assume here no torsion). In order to preserve the covariance of the
tetrad mapping we must have 
\begin{eqnarray}
d_\mu e_{\nu,A} =  \partial_\mu e_{A,\nu} - \Gamma_{\nu
\mu}^\lambda e_{A,\lambda} + \omega_{AB \mu } e_\nu^B =0  .
\end{eqnarray} 
In addition, the condition $d_\mu g^{\mu \nu} = 0 $, implying  
\begin{eqnarray}
d_\mu \eta_{AB}= \omega_{AB\mu}  + \omega_{BA\mu} =0,  
\end{eqnarray} 
requires an antisymmetric spin connection, $ \omega_{AB\mu}=
-\omega_{BA\mu}$, given by
\begin{eqnarray}
\omega_{AB\mu} = e_A^\nu \left[ \partial_\mu e_{B,\nu} - \Gamma_{\nu
\mu}^\lambda e_{B,\lambda} \right] .
\end{eqnarray} 
The frame and coordinate covariant derivative $d_\mu $ is defined
according to the spin of the corresponding field. For a spin-0 $U$,
spin-1/2, $\Psi$, spin-1, $ A_\mu $, and spin 3/2, $ \Psi_\mu $, fields
the transformation properties are
\begin{eqnarray}
U(x) &\to & U(x), \nonumber \\ \Psi (x) &\to& S(\Lambda(x)) \Psi (x), \\
A_\mu (x) &\to & \frac{\partial x'_\nu}{\partial x^\mu} A_\nu (x), \\
\Psi_\mu (x) &\to & \frac{\partial x'_\nu}{\partial x^\mu} S ( \Lambda
(x) ) \Psi_\nu (x).
\end{eqnarray} 
For infinitesimal Lorentz transformations $\Lambda^A_B = \delta^A_B +
\epsilon^A_B$ with $ \epsilon_{AB} = - \epsilon_{BA} $ one has $
S(\Lambda) = 1 -\frac{i}4 \sigma_{AB} \epsilon^{AB} $ with $\sigma_{AB}$
defined below (see Eq.~(\ref{eq:sigma})).  

For a scalar (spin-0) field we have the standard definition
\begin{eqnarray}
d_\mu U = \partial_\mu U .
\end{eqnarray} 
For a vector (spin-1), one has 
\begin{eqnarray}
d_\mu A_\nu = \partial_\mu A_{\nu} - \Gamma_{\nu \mu}^\lambda A_\lambda ,
\end{eqnarray} 
fulfilling the property 
\begin{eqnarray}
\left[ d_\mu , d_\nu \right] A_\alpha = R^\lambda_{\, \,  \alpha \mu \nu }
A_\lambda  
\label{eq:dmudnuA} 
\end{eqnarray}
with the Riemann curvature tensor given by Eq.~(\ref{eq:curvature}).
The coordinate and Lorentz covariant derivative for Dirac fermions
(spin 1/2) is defined as
\begin{eqnarray}
d_\mu \Psi = \partial_\mu \Psi(x) - i \omega_\mu \Psi(x) ,
\end{eqnarray} 
where $\omega_\mu$ is the Cartan spin connection,
\begin{eqnarray} 
\omega_\mu = \frac14 \sigma^{AB} \omega_{AB\mu},
\end{eqnarray} 
and
\begin{eqnarray}
\sigma_{AB} = \frac{i}2 [ \gamma_A , \gamma_B ] , 
\label{eq:sigma} 
\end{eqnarray} 
with the $\gamma_A$ are fixed $x-$independent Dirac matrices (we use
the conventions of Ref.~\cite{IZ80}) fulfilling the standard flat
space anticommutation rules, 
\begin{eqnarray}
\gamma^A \gamma^B  + \gamma^B  \gamma^A  = 2
\eta^{AB}   .
\end{eqnarray}
The space-time dependent Dirac matrices are defined as
\begin{eqnarray}
\gamma_\mu (x) = \gamma_A e^A_\mu (x)
\end{eqnarray} 
and fulfill 
\begin{eqnarray}
\gamma^\mu (x) \gamma^\nu (x) + \gamma^\nu (x) \gamma^\mu (x) = 2
g^{\mu \nu} (x)  .
\end{eqnarray} 
The covariant derivative of a
frame ($x$-independent) Dirac matrix (behaving as the adjoint
representation $ \Psi \bar \Psi $) is
\begin{eqnarray}
d_\mu \gamma_A = \partial_\mu \gamma_A - i \left[ \omega_\mu ,
\gamma_A \right] + \omega_{AB\mu} \gamma^B =0 .   
\end{eqnarray} 
Thus, we obtain the useful identity for the coordinate (and $x$-dependent)
Dirac matrix,
\begin{eqnarray}
d_\mu \gamma_\nu (x) =0,  
\end{eqnarray} 
which implies that for the free Dirac operator the order is irrelevant
$\slashchar{d} \Psi =\gamma^\mu (x) d_\mu \Psi =d_\mu \gamma^\mu (x)
\Psi $.  For a mixed tensor (spin-3/2) the frame and coordinate
covariant derivative reads
\begin{eqnarray}
d_\nu \Psi_\mu = \Psi_{\nu; \mu} = \partial_\mu \Psi_{\nu} - \Gamma_{\nu
\mu}^\lambda \Psi_\lambda - i \omega_\nu \Psi_\mu .
\end{eqnarray} 
Applying the previous definition to $d_\mu \Psi $ one gets the useful
formulas
\begin{eqnarray}
\left[ d_\mu , d_\nu \right] \Psi &=& + \frac{i}4 \sigma^{\alpha
\beta} R_{\alpha \beta \mu \nu} \Psi, \\ d^\mu d_\mu \Psi &=&
\frac{1}{\sqrt{-g}} \left\{ \left(\partial_\mu-i \omega_\mu \right)
\left[ \sqrt{-g} g^{\mu \nu} \left(\partial_\nu-i \omega_\nu \right)
\right] \Psi \right\}, \nonumber \\ 
\end{eqnarray}
where $ \sigma^{\alpha \beta } = e^\alpha_A e^\beta_B \sigma^{AB} $ is
an antisymmetric x-dependent matrix. 

Gauge fields can be included by the standard minimal substitution rule
yielding the covariant derivative for a
fermion, 
\begin{eqnarray}
\nabla_\mu \Psi = \left( d_\mu - i A_\mu \right) \Psi .
\end{eqnarray} 
With this notation the full Dirac operator in the presence of external
vector, axial-vector, scalar, pseudoscalar and gravitational fields
reads as in Eq.~(\ref{eq:dirac_op}), where 
\begin{eqnarray}
\slashchar{A}  = \gamma^\mu (x) A_\mu (x) ,
\end{eqnarray} 
and the pseudoscalar Dirac matrix in the curved case is defined as 
\begin{eqnarray}
\gamma_5 (x) &=& \frac{1}{4! \sqrt{-g}} \epsilon^{\mu\nu\alpha\beta}
\gamma_\mu (x) \gamma_\nu (x) \gamma_\alpha (x) \gamma_\beta (x)
\nonumber \\ &=& \frac{1}{4!} \epsilon^{ABCD} \gamma_A \gamma_B
\gamma_C \gamma_D = \gamma_5 .
\end{eqnarray} 
Here $ g(x) = {\rm det} ( g^{\mu \nu} ) $ since $ {\rm det} ( e_A^\nu
)^2 = {\rm det} ( g^{\mu \nu} ) $ with $ \epsilon^{0123} = 1 $ (both
in the frame as well as in the coordinate sense).

The full coordinate, frame and chiral gauge covariant derivative for
pseudoscalar (spin-0), Dirac spinor (spin-1/2) and a Rarita-Schwinger
spinor (spin 3/2) fields are given by the following formulas,
\begin{eqnarray}
\nabla_\mu U &=& {\cal D}_\mu U = \partial_\mu U - i [ v_\mu ,U] - i
\{ a_\mu ,U \}, \nonumber \\ \nabla_\mu \Psi &=& {\cal D}_\mu \Psi =
\partial_\mu \Psi - i ( \omega _\mu + v_\mu + \gamma_5 a_\mu ) \Psi,
\nonumber \\ \nabla_\mu \Psi_\nu &=& \partial_\mu \Psi_\nu - i(
\omega_\mu+v_\mu + \gamma_5 a_\mu ) \Psi_\nu - \Gamma_{\nu
\mu}^\lambda \Psi_\lambda , \nonumber \\
\end{eqnarray} 
and they correspond to replacing the derivative by the frame and
coordinate covariant derivative, $\partial_\mu \to d_\mu $, in the
chiral covariant derivative $D_\mu $. Note that with this definition
neither $ {\cal D}_\mu {\cal D}_\nu \Psi \neq \nabla_\mu \nabla_\nu
\Psi $ nor $ {\cal D}_\mu {\cal D}_\nu U $ are coordinate covariant
since the second derivative does not include the Riemann connection
$\Gamma^\lambda_{\mu\nu}$.

\subsection{The Second order Operator} 

In the absence of gravitational sources, the normal parity
contribution can be obtained from the second order operator (see Eq.~(\ref{eq:np})), 
\begin{eqnarray}
{\bf D}_5 {\bf D} &=& \left[ \slashchar{D}_L^2 + {\rm i} {\cal
M}^\dagger 
\slashchar{D}_L - {\rm i} \slashchar{D}_R {\cal M}^\dagger  + {\cal M}^\dagger {\cal
M} \right] P_R \nonumber \\  &+& 
\left[ \slashchar{D}_R^2 + {\rm i} {\cal M}
\slashchar{D}_L - {\rm i} \slashchar{D}_R {\cal M} + {\cal M} {\cal
M}^\dagger \right] P_L . \nonumber  \\ 
\label{eq:KG-flat} 
\end{eqnarray} 
Gravitational fields can be coupled by covariantizing first the Dirac
operator, i.e. making $ \partial_\mu \to d_\mu $ or $ D_\mu \to {\cal
D}_\mu $ and taking into account that since a spinor field is a
coordinate scalar we have 
\begin{eqnarray}
{\cal D}_\mu \Psi = \nabla_\mu \Psi .  
\end{eqnarray} 
The same reasoning can be applied to the coordinate scalar
$\slashchar{\nabla} \Psi $, yielding 
\begin{eqnarray}
{\cal D}_\mu \slashchar{\nabla} \Psi = \nabla_\mu \slashchar{\nabla} \Psi  .
\end{eqnarray} 
This means that we can assume $ \slashchar{{\cal D}}_{L,R} =
\slashchar{\nabla}_{L,R} $ when acting on spinor field as follows  
\begin{eqnarray} 
{\bf D}_5 {\bf D} \Psi &=& \left[ \slashchar{\nabla}_L^2 + {\rm i}
{\cal M} \slashchar{\nabla}_L - {\rm i} \slashchar{\nabla}_R {\cal M}
+ {\cal M}^\dagger {\cal M} \right] P_R \Psi \nonumber \\ &+& \left[
\slashchar{\nabla}_R^2 + {\rm i} {\cal M}^\dagger \slashchar{\nabla}_L
- {\rm i} \slashchar{\nabla}_R {\cal M}^\dagger + {\cal M} {\cal
M}^\dagger \right] P_L \Psi .\nonumber \\
\label{eq:KG-curved} 
\end{eqnarray} 
If we include the gauge fields we have two vector like theories with
left and right gauge fields $V_\mu^L$ and $V_\mu^R$
respectively. Suppressing momentarily the left and right labels we
have   
\begin{eqnarray}
\slashchar{{\cal D}}^2 \Psi = \slashchar{{\nabla}}^2 \Psi = \left[
{\nabla}^\mu {\cal \nabla}_\mu - \frac12 \sigma^{\mu \nu} F_{\mu \nu}
+ \frac14 R \right] \Psi , 
\end{eqnarray} 
where the use of the identity 
\begin{eqnarray}
\left[ \nabla_\mu , \nabla_\nu \right] \Psi &=& \left[ {\cal D}_\mu ,
{\cal D}_\nu \right] \Psi \nonumber \\ &=& \left[ D_\mu , D_\nu
\right] \Psi + \frac{i}4 \sigma^{\alpha \beta} R_{\alpha \beta \mu
\nu} \Psi 
\end{eqnarray} 
has been made. The coordinate and frame invariant Laplacian for a
Dirac spinor is given by
\begin{eqnarray}
\nabla^\mu \nabla_\mu \Psi =  \frac{1}{\sqrt{-g}} {\cal D}_\mu \left( 
\sqrt{-g} g^{\mu \nu} {\cal D}_\nu \Psi \right) .
\end{eqnarray} 
Note that for a Dirac spinor field $\Psi $ the operator ${\cal D}_\mu
$ contains the spin connection. Reinserting the right and left chiral
notation the second order operator takes the suitable form
\begin{eqnarray}
{\bf D}_5 {\bf D} &=& \frac{1}{\sqrt{-g}} \left[ {\cal D}_\mu \left( 
\sqrt{-g} g^{\mu \nu} {\cal D}_\nu \right) \right] + {\cal V},
\label{eq:KG-heat} 
\end{eqnarray} 
with 
\begin{eqnarray}
{\cal V}=  {\cal V}_R P_R + {\cal V}_L  P_L   
\end{eqnarray} 
and 
\begin{eqnarray}
{\cal V}_R &=& -\frac12 \sigma^{\mu \nu} {F}_{\mu \nu}^R  
+ \frac14 R - {\rm i} \gamma^\mu {\nabla}_\mu  {\cal M} + {\cal
M}^\dagger {\cal M} ,\nonumber \\ \\ 
{\cal V}_L &=& -\frac12 \sigma^{\mu \nu} {F}_{\mu \nu}^L  
+ \frac14 R - {\rm i} \gamma^\mu {\nabla}_\mu  {\cal M}^\dagger + {\cal
M} {\cal M}^\dagger .   \nonumber 
\end{eqnarray} 

\subsection{Derivative expansion} 

We use the proper-time representation,
\begin{eqnarray}
{\rm Tr} \log \left( {\bf D}_5 {\bf D} \right) = - {\rm Tr} \int_0^\infty
\frac{d \tau}{\tau} e^{- {\rm i}  \tau {\bf D}_5 {\bf D} } + C ,
\end{eqnarray} 
with $C$ and infinite constant. The form of the operator $ {\bf D}_5
{\bf D} $ in Eq.~(\ref{eq:KG-heat}) is suitable to make a heat kernel
expansion in curved space-time as the one of
Ref.~\cite{Luscher:1982wf}. For a review see
e.g. \cite{Vassilevich:2003xt} and references therein. In our
particular case, before undertaking the heat kernel expansion we
separate a $\omega^2$ contribution from the operator ${\bf D}_5 {\bf
D}$ which we treat exactly,
\begin{eqnarray}
\langle x | e^{-{\rm i} \tau {\bf D}_5 {\bf D} } | x \rangle &=& e^{ -
{\rm i}\tau \omega^2 } \langle x | e^{-{\rm i} \tau ( {\bf D}_5 {\bf
D} - \omega^2 ) } | x \rangle \\ &=& \frac{\rm i}{(4\pi {\rm
i}\tau )^2} e^{-{\rm i} \tau \omega^2 } \sum_{n=0}^\infty a_{2n} (x)
\left( {\rm i} \tau \right)^n . \nonumber 
\end{eqnarray}
The derivative expansion is done by considering $U$ zeroth order the
vector and axial fields $ v_\mu $ and $a_\mu$ first order, and any
derivative $\partial_\mu $ first order. This implies in particular
that $R^{\mu\nu\alpha\beta} $, $R^{\mu \nu } $ and $R$ are taken to be
of second order. Finally, the external scalar and pseudoscalar fields
$s$ and $p$ are taken to be second order as well. Thus, the
multiplicative operator ${\cal V}-\w^2 $ is at least first order in the
chiral counting. To the computed order $ {\cal O} (p^4) $ in the heat
kernel expansion one has to go up to $ a_4 $. The contributions can be
separated into the flat space non-vanishing contributions and the
curvature contributions generated by quantum effects. Using the form
suggested in \cite{Parker:dj} we have 
\begin{eqnarray}
a_0 &=& 1, \nonumber \\ 
a_1 &=& \omega^2-{\cal V} + \frac16 R ,\nonumber \\ 
a_2 &=& \frac1{180} R_{\mu \nu \alpha \beta} R^{\mu \nu \alpha \beta}-
\frac1{180} R_{\mu \nu } R^{\mu \nu} + \frac{1}{12} {\cal F}^{\mu \nu}
{\cal F} _{\mu \nu} \nonumber \\ &+& \frac1{30} \nabla^2 R - \frac16
\nabla^2 {\cal V}  + \frac12 \left[ \omega^2-{\cal V} + \frac16 R
\right]^2  ,
\nonumber \\
a_3 &=& \frac16 \left[ \omega^2-{\cal V} + \frac16 R
\right]^3  - \frac1{12} \nabla^\mu {\cal V} \nabla_\mu
{\cal V} + {\cal O} (p^6) ,\nonumber \\
a_4  &=& \frac1{24}\left[{\cal V}-\omega^2\right]^4   + {\cal O} (p^6) ,
\end{eqnarray} 
where 
\begin{eqnarray}
{\cal F}_{\mu \nu} &=& i \left[ {\cal D}_\mu , {\cal D}_\nu \right ],
\\ \nabla^2 {\cal V} &=& \nabla^\mu \nabla_\mu {\cal V}.
\end{eqnarray} 
Clearly, the heat kernel coefficients depend on the spectral mass $\w$
in a polynomial fashion. Using the integrals
\begin{eqnarray}
\int_0^\infty \frac{d\tau}{\rm \tau} ({\rm i} \tau)^{z-2} e^{-{\rm i}
\tau \w^2} =(\w^2)^z \Gamma(z-2)
\end{eqnarray} 
we get for integer $z=n$ and after using the spectral conditions,
Eq.~({\ref{rhon}), the normal parity contribution of the action takes
the form
\begin{eqnarray}
-\frac{\rm i}2 {\rm Tr} \log {\bf D}_5 {\bf D} &=& -\frac12
 \frac{N_c}{(4\pi^ 2)} \int d^4 x \sqrt{-g} \int d\w \rho(\w)
 \nonumber \\ &\times& {\rm tr} \langle -\frac12 \w^4 \log \w^2 a_0 +
 \w^2 \log \w^2 a_1 \nonumber \\ &-&\log (\w^2/\mu^2) a_2 +
 \frac1{\w^2} a_3 + \frac1{\w^4} a_4 + \dots \rangle \nonumber \\
&=& \int d^4 x \sqrt{-g} \left({\cal L}^{(2)} + {\cal L}^{(4)} + \dots
 \right) . \nonumber \\
\end{eqnarray} 
After evaluation of the Dirac traces, the second order Lagrangean is
\begin{eqnarray}
{\cal L}^{(2)} &=&{N_c\over(4\pi)^2} \int \rho(\omega) \Big\{ -
\omega^2 \log \omega^2 \langle \nabla_\mu U^\dagger \nabla^\mu U
\rangle \nonumber \\ &+& 2 \omega^3 \log \omega^2 \langle {\it
m}^\dagger U + U^\dagger {\it m} \rangle + \omega^2 \log \omega^2
\frac1{12} \langle R \rangle \Big\}, \nonumber \\
\end{eqnarray} 
whereas the fourth order becomes
\begin{eqnarray} 
{\cal L}^{(4)} &=& {N_c\over (4\pi)^2} \int \rho(\omega) \Big\{
\nonumber \\
&+& {1\over6} \log \omega^2 \langle (F_{\mu\nu}^R)^2 +
(F_{\mu\nu}^L)^2 \rangle \nonumber \\ &-& \log \omega^2 \langle
\frac7{720} R^{\alpha \beta \mu \nu} R_{\alpha \beta \mu \nu} +
\frac{1}{144} R^2 + \frac1{90} R^{\mu \nu} R_{\mu \nu} \rangle
\nonumber \\
&-&\frac{i}3 \langle F_{\mu\nu}^R \nabla_\mu U^\dagger \nabla_\nu U +
F_{\mu\nu}^L \nabla_\mu U \nabla_\nu U^\dagger \rangle \nonumber \\ 
&+& \frac1{12} \langle (\nabla_\mu U \nabla_\nu U^\dagger )^2 \rangle
-\frac16 \langle (\nabla_\mu U \nabla^\mu U^\dagger )^2 \rangle
\nonumber \\
&+& \frac16 \langle \nabla^\mu \nabla^\nu U \nabla_\mu \nabla_\nu 
U^\dagger \rangle- \frac16 \langle F_{\mu\nu}^L U F_{\mu\nu}^R
U^\dagger \rangle \nonumber \\ 
&+&  \log \omega^2 \omega^2 \left( 2 \langle {\it m}^\dagger{\it m} \rangle
+ \langle ( {\it m}^\dagger
U+U^\dagger {\it m} )^2 \rangle \right) 
\nonumber \\
&-& {1\over2} \omega \langle \nabla_\mu U^\dagger \nabla^\mu U ({\it m}^\dagger
U+U^\dagger {\it m} ) \rangle \nonumber \\
&-&\log \omega^2 \omega  \langle \nabla_\mu U^\dagger \nabla^\mu {\it m} +
\nabla_\mu {\it m}^\dagger \nabla^\mu U \rangle \nonumber \\ 
&-&\omega \log \omega^2 \frac16 R \langle U^\dagger {\it m}+ {\it
m}^\dagger U \rangle + \frac1{12} R\, \nabla_\mu U^\dagger \nabla^\mu U
\rangle  \Big\} . \nonumber \\ 
\end{eqnarray} 
Note that up to this order the moments $ \rho_0 =1 $, $\rho_1=0$ and
$\rho_2=0$ as well as the log-moments $ \rho^\prime_0 $,
$\rho^\prime_1 $ and $\rho^\prime_2 $ appear. 

\subsection{Equations of Motion} 

We define,
\begin{eqnarray} 
\chi = 2 B_0 {\it m} = 2 B_0 \left( s + i p \right).
\end{eqnarray} 
For on-shell pseudoscalars one may minimize the action at lowest order
\begin{eqnarray}
S^{(2)} &=& {f^2\over 4} \int d^4 x \sqrt{-g} \nonumber \\ &\times&
\langle \nabla_\mu U^\dagger \nabla^\mu U +(\chi^\dagger U + U^\dagger
\chi) - \frac1{12} R \rangle , \nonumber \\
\end{eqnarray} 
to obtain the equations of motion. Since $U$ is unitary, $ U^\dagger U
= 1$, we have that the variations on $ U$ and $ U^\dagger $ are not
independent of each other, $ \delta U^\dagger U + U^\dagger \delta U=0
$. For SU(3)-flavour one has, in addition, to impose the condition $
\Det U=1$.  One can treat $U$ and $U^\dagger$ independently by
introducing a term in the Lagrangian of the form $ \langle \Lambda
U^\dagger U - i \lambda \log U \rangle $ where the Lagrange
multipliers are $\Lambda $, a hermitean matrix, and $\lambda$, a real
c-number. Thus, the EOM are
\begin{eqnarray}
\nabla^2 U &=& \chi +  (\Lambda-i \lambda) U , \nonumber \\ 
\nabla^2 U^\dagger  &=& \chi^\dagger  + U^\dagger ( \Lambda + i \lambda) ,
\end{eqnarray} 
where 
\begin{eqnarray}
\nabla^2 U =  \frac{1}{\sqrt{-g}} D_\mu \left( 
\sqrt{-g} g^{\mu \nu} D_\nu U \right)  .
\end{eqnarray} 
Combining these two equations, we get
\begin{eqnarray}
U^\dagger \nabla^2 U - \nabla^2 U^\dagger U &=& U^\dagger \chi
-\chi^\dagger U - 2 i \lambda .
\end{eqnarray}
Taking the trace and using the condition that for a matrix with $\Det
U=1$ one has $ \langle U^\dagger \nabla_\mu U \rangle=0$ and hence $
\langle U^\dagger \nabla^2 U - \nabla^2 U^\dagger U \rangle =0 $, we
get
\begin{eqnarray}
\lambda = \frac1{6i} \langle U^\dagger \chi - \chi^\dagger U \rangle ,   
\end{eqnarray} 
thus
\begin{eqnarray}
U^\dagger \nabla^2 U - \nabla^2 U^\dagger U &=& U^\dagger \chi -\chi^\dagger U -
\frac13 \langle U^\dagger \chi -\chi^\dagger U  \rangle . \nonumber \\ 
\end{eqnarray}
On the other hand $ \Lambda $ is given by
\begin{eqnarray} 
2 \Lambda = \nabla^2 U^\dagger U + U \nabla^2 U^\dagger - ( \chi U^\dagger +
\chi^\dagger U ) .
\label{eq:aa}
\end{eqnarray}
Using the identities deduced form the unitarity condition $U^\dagger
U=1$,
\begin{eqnarray}
U^\dagger \nabla_\mu U + \nabla_\mu U^\dagger U &=& 0 \\ U^\dagger
\nabla^2 U + \nabla^2 U^\dagger U &=& - 2 \nabla_\mu U^\dagger
\nabla^\mu U,
\label{eq:bb}
\end{eqnarray} 
and combining them with the previous Eqs.~(\ref{eq:aa},\ref{eq:bb}) we
get the identities
\begin{eqnarray} 
\langle \nabla^2 U^\dagger \nabla^2 U \rangle &=&  \langle \left(
\nabla_\mu U^\dagger \nabla^\mu U \right)^2 \rangle 
-\frac14 \langle \left(
\chi^\dagger U - U^\dagger \chi \right)^2 \rangle \nonumber \\ 
&+&\frac1{12} \langle 
\chi^\dagger U - U^\dagger \chi  \rangle^2 
\label{eq:id1}
\end{eqnarray} 
and 
\begin{eqnarray} 
\langle \chi^\dagger \nabla^2 U + \nabla^2 U^\dagger \chi \rangle &=& 2 \langle
\chi^\dagger \chi \rangle - \frac12 \langle \left( \chi^\dagger U +
U^\dagger \chi \right)^2 \rangle \nonumber \\ &-& \langle \left(
\chi^\dagger U + U^\dagger \chi \right) \nabla^\mu U^\dagger \nabla_\mu U
\rangle \nonumber \\ 
&+& \frac16 \langle \chi^\dagger U +
U^\dagger \chi  \rangle^2 .
\label{eq:id2}
\end{eqnarray} 
In the case of the $U(3)$ group one has $\Det U = e^{i \eta_0 /f} \neq
1 $ and the last two terms involving $ \langle \chi^\dagger U \pm
U^\dagger \chi \rangle^2 $ in Eqs.~(\ref{eq:id1}) and (\ref{eq:id2})
should be dropped. (See the discussion before Eq.~(\ref{eq:lec's_g}))
The result can be further simplified using the integral identity
\begin{eqnarray}
&& \int d^4 x \sqrt{-g} \, \langle \nabla^\mu \nabla^\nu U^\dagger
\nabla^\mu \nabla^\nu U \rangle = \int d^4 x \sqrt{-g} \langle
\nabla^2 U^\dagger \nabla^2 U \rangle \nonumber \\ && + \int d^4 x
\sqrt{-g} R_{\mu \nu} \langle \nabla^\mu U^\dagger \nabla^\nu U
\rangle ,
\label{eq:ricci}
\end{eqnarray} 
which can be deduced from Eq.~(\ref{eq:dmudnuA}) applied to
$\nabla_\mu U $. Finally, we also have the SU(3) identity
\begin{eqnarray}
\langle (\nabla_\mu U^\dagger \nabla_\nu U)^2 \rangle &=& - 2 \langle
\nabla_\mu U^\dagger \nabla^\mu U \rangle^2 \label{eq:su3} \\ &+& \langle \nabla_\mu
U^\dagger \nabla_\nu U \rangle^2 + \frac12 \langle \nabla_\mu
U^\dagger \nabla^\mu U \rangle^2 \nonumber .
\end{eqnarray} 
Once the identities (\ref{eq:id1}),(\ref{eq:id2}),(\ref{eq:ricci}) and
(\ref{eq:su3}) have been used one can make the substitute the
coordinate-frame-covariant derivative by the covariant derivative,
i.e., $ \nabla^\mu U = D^\mu U $, since the pseudoscalar matrix $U$ is
a coordinate and frame scalar. In that way Eqs.~(\ref{eq:chl4}) and
(\ref{eq:chl4R}) are deduced.

In four dimensions, one can
reduce the form of the curvature contributions to the Lagrangean if
the Gauss-Bonnet theorem is used in Eq.~(\ref{eq:chl4R}), namely that  
\begin{eqnarray} 
\kappa = \int d^4 x \sqrt{-g} \left[ R^2 - 4 R_{\mu \nu} R^{\mu \nu} + 
R_{\mu \nu \alpha \beta} R^{\mu \nu \alpha \beta} \right]
\end{eqnarray} 
is a topological invariant (the Euler number ) and hence 
\begin{eqnarray}
\delta \kappa =0  
\end{eqnarray} 
under metric deformations, $ g_{\mu \nu} \to g_{\mu\nu} + \delta
g_{\mu\nu}$. This relation was not taken into account in
Ref.~\cite{Donoghue:qv} but it does not affect the calculation of the
energy momentum tensor in flat space, Eq.~(\ref{eq:en-mom}). 

\subsection{Derivative expansion for first order differential operators} 

As we see the definition of the action involves the Dirac operator
${\bf D}$ only, which is a first order differential operator. The
derivative expansion of the Dirac operator can be done using the
identity 
\begin{eqnarray}
 \langle x | \frac1{i \slashchar{D} - {\cal M} - \omega U } | x
\rangle = \int \frac{d^4 k}{(2 \pi)^4 } \frac1{\slashchar{k} + i
\slashchar{D} - {\cal M} - \omega U }, \nonumber \\  
\label{eq:bas1}
\end{eqnarray} 
where the differential operator acts on the right. This formula can be
justified by requiring vector gauge invariance of the
action~\cite{Chan:1986jq} or by using the asymmetric version of the
Wigner transformation presented in
Ref.~\cite{Salcedo:1994qy}. Expanding in powers of $D$ and ${\cal M}$
and squaring the denominator we get
\begin{eqnarray} 
 \langle x | \frac1{i \slashchar{D} - {\cal M} - \omega U } | x
\rangle 
&=&  \sum_{n=0}^\infty \int \frac{d^4 k}{(2 \pi)^4 }
\left[\frac{-1}{k^2-\omega^2} \right]^{n+1}  \nonumber \\ && \!\!\!\!\!\!\!\!\!\!\!\!\left(\slashchar{k} +
\omega U^\dagger \right) \left[ \left(i \slashchar{D} - {\cal M}
\right) \left(\slashchar{k} + \omega U^\dagger \right) \right]^n . \nonumber \\
\label{eq:bas2}
\end{eqnarray} 
In this way Eq.~(\ref{eq:wzw}) can be derived.



\begin{thebibliography}{66}

\expandafter\ifx\csname natexlab\endcsname\relax\def\natexlab#1{#1}\fi
\expandafter\ifx\csname bibfnamefont\endcsname\relax
  \def\bibfnamefont#1{#1}\fi
\expandafter\ifx\csname citenamefont\endcsname\relax
  \def\citenamefont#1{#1}\fi
\expandafter\ifx\csname url\endcsname\relax
  \def\url#1{\texttt{#1}}\fi
\expandafter\ifx\csname urlprefix\endcsname\relax\def\urlprefix{URL }\fi
\providecommand{\bibinfo}[2]{#2}
\providecommand{\eprint}[2][]{\url{#2}}





\bibitem{Langacker:1973hh}
P.~Langacker and H.~Pagels,
Phys.\ Rev.\ D {\bf 8} (1973) 4595.

\bibitem{Weinberg:1978kz}
S.~Weinberg,
Physica A {\bf 96} (1979) 327.


\bibitem{Gasser:1983yg}
J.~Gasser and H.~Leutwyler,
Annals Phys.\  {\bf 158} (1984) 142.

\bibitem{Gasser:1984gg}
J.~Gasser and H.~Leutwyler,
Nucl.\ Phys.\ B {\bf 250} (1985) 465.

\bibitem{Donoghue:qv}
J.~F.~Donoghue and H.~Leutwyler,
Z.\ Phys.\ C {\bf 52} (1991) 343.



\bibitem{Pich:1995bw}
A.~Pich,
Rept.\ Prog.\ Phys.\  {\bf 58}, 563 (1995)

\bibitem{Bijnens:1998fm}
J.~Bijnens, G.~Colangelo and P.~Talavera,
JHEP {\bf 9805}, 014 (1998)
\bibitem{Colangelo:2001df}
G.~Colangelo, J.~Gasser and H.~Leutwyler,
Nucl.\ Phys.\ B {\bf 603}, 125 (2001)
\bibitem{Amoros:2001cp}
G.~Amoros, J.~Bijnens and P.~Talavera,
Nucl.\ Phys.\ B {\bf 602}, 87 (2001)
\bibitem{Bijnens:2002hp}
J.~Bijnens and P.~Talavera,
JHEP {\bf 0203} (2002) 046

\bibitem{Ananthanarayan:2000ht}
B.~Ananthanarayan, G.~Colangelo, J.~Gasser and H.~Leutwyler,
Phys.\ Rept.\  {\bf 353} (2001) 207

\bibitem{Yndurain:2002ud}
F.~J.~Yndurain,


\bibitem{Pelaez:2003eh}
J.~R.~Pelaez and F.~J.~Yndurain,
Phys.\ Rev.\ D {\bf 68} (2003) 074005


\bibitem{Ecker:1988te}
G.~Ecker, J.~Gasser, A.~Pich and E.~de Rafael,
Nucl.\ Phys.\ B {\bf 321} (1989) 311.

\bibitem{Pich:2002xy}
A.~Pich,
arXiv:hep-ph/0205030.

\bibitem{Kubis:1999db}
B.~Kubis and U.~G.~Meissner,
Nucl.\ Phys.\ A {\bf 671}, 332 (2000)
[Erratum-ibid.\ A {\bf 692}, 647 (2001)]


\bibitem{Diakonov:tw}
D.~Diakonov and M.~I.~Eides,
JETP Lett.\  {\bf 38}, 433 (1983)
[Pisma Zh.\ Eksp.\ Teor.\ Fiz.\  {\bf 38}, 358 (1983)].
\bibitem{Balog:ps}
J.~Balog,
Phys.\ Lett.\ B {\bf 149}, 197 (1984).
\bibitem{Andrianov:ay}
A.~A.~Andrianov,
Phys.\ Lett.\ B {\bf 157}, 425 (1985).
\bibitem{Belkov:mb}
A.~A.~Belkov, D.~Ebert and V.~N.~Pervushin,
Phys.\ Lett.\ B {\bf 193} (1987) 315.
\bibitem{Espriu:1989ff}
D.~Espriu, E.~de Rafael and J.~Taron,
Nucl.\ Phys.\ B {\bf 345} (1990) 22
[Erratum-ibid.\ B {\bf 355} (1991) 278].
\bibitem{Hansson:jy}
T.~H.~Hansson, M.~Prakash and I.~Zahed,
Nucl.\ Phys.\ B {\bf 335} (1990) 67.
\bibitem{Holdom:iq}
B.~Holdom, J.~Terning, and K.~Verbeek,
Phys.\ Lett.\ B {\bf 245} (1990) 612.
\bibitem{RuizArriola:gc}
E.~Ruiz Arriola,
Phys.\ Lett.\ B {\bf 253} (1991) 430.
\bibitem{Bernard:1991wy}
V.~Bernard and U.~G.~Meissner,
Phys.\ Lett.\ B {\bf 266} (1991) 403.
\bibitem{Schuren:1991sc}
C.~Schuren, E.~Ruiz Arriola and K.~Goeke,
Nucl.\ Phys.\ A {\bf 547}, 612 (1992).
\bibitem{Bijnens:1992uz}
J.~Bijnens, C.~Bruno and E.~de Rafael,
Nucl.\ Phys.\ B {\bf 390} (1993) 501
\bibitem{Schuren:1993aj}
C.~Schuren, F.~Doring, E.~Ruiz Arriola and K.~Goeke,
Nucl.\ Phys.\ A {\bf 565} (1993) 687.
\bibitem{Polyakov:1995vh}
M.~V.~Polyakov and V.~V.~Vereshagin,
Phys.\ Rev.\ D {\bf 54} (1996) 1112
\bibitem{Wang:1999cp}
Q.~Wang, Y.~P.~Kuang, M.~Xiao and X.~L.~Wang,
Phys.\ Rev.\ D {\bf 61} (2000) 054011

\bibitem{RuizArriola:2002wr}
E.~Ruiz Arriola,
Acta Phys.\ Polon.\ B {\bf 33}, 4443 (2002)

\bibitem{Sutton:1991ay}
P.~J.~Sutton, A.~D.~Martin, R.~G.~Roberts and W.~J.~Stirling,
Phys.\ Rev.\ D {\bf 45} (1992) 2349.

\bibitem{Davidson:1994uv}
R.~M.~Davidson and E.~Ruiz Arriola,
Phys.\ Lett.\ B {\bf 348}, 163 (1995).


\bibitem{Davidson:2001cc}
R.~M.~Davidson and E.~Ruiz Arriola,
Phys.\ Lett.\ B {\bf 348}, 163 (1995). 
Act. Phys.\ Pol.\ B {\bf 33}, 1791 (2002)

\bibitem{Broniowski:2003rp}
W.~Broniowski and E.~Ruiz Arriola,
hep-ph/0307198.

\bibitem{e615} J. S. Conway {\em et al.}, Phys. Rev. D {\bf 39} (1989) 92.

\bibitem{RuizArriola:2002bp}
E.~Ruiz Arriola and W.~Broniowski,
Phys.\ Rev.\ D {\bf 66}, 094016 (2002)

\bibitem{Dalley:2002nj}
S.~Dalley and B.~van de Sande,
Phys.\ Rev.\ D {\bf 67} (2003) 114507.



\bibitem{Dalley:2003sz}
S.~Dalley,
Phys.\ Lett.\ B {\bf 570}, 191 (2003)
[arXiv:hep-ph/0306121].

\bibitem{RuizArriola:2001rr}
E.~Ruiz Arriola,
hep-ph/0107087.

\bibitem{RuizArriola:2003bs}
E.~Ruiz Arriola and W.~Broniowski,
Phys.\ Rev.\ D {\bf 67}, 074021 (2003)

\bibitem{Wess:yu}
J.~Wess and B.~Zumino,
Phys.\ Lett.\ B {\bf 37} (1971) 95.


\bibitem{Witten:tw}
E.~Witten,
Nucl.\ Phys.\ B {\bf 223} (1983) 422.

\bibitem{Bowman:2002bm}
P.~O.~Bowman, U.~M.~Heller and A.~G.~Williams,
Phys.\ Rev.\ D {\bf 66} (2002) 014505

\bibitem{Efimov:1988yd}
G.~V.~Efimov and M.~A.~Ivanov,
Int.\ J.\ Mod.\ Phys.\ A {\bf 4} (1989) 2031.

\bibitem{Broniowski:2003wn}
W.~Broniowski and E.~Ruiz Arriola,
arXiv:hep-ph/0310048.

\bibitem{RuizArriola:2003wi}
E.~Ruiz Arriola and W.~Broniowski,
arXiv:hep-ph/0310044.

\bibitem{Delbourgo:1977jc}
R.~Delbourgo and P.~C.~West,
J.\ Phys.\ A {\bf 10} (1977) 1049.

\bibitem{Ripka:am}
G.~Ripka and S.~Kahana,
Phys.\ Lett.\ B {\bf 155} (1985) 327.

\bibitem{Soni:fy}
V.~Soni,
Phys.\ Lett.\ B {\bf 183} (1987) 91.

\bibitem{Pagels:hd}
H.~Pagels and S.~Stokar,
Phys.\ Rev.\ D {\bf 20} (1979) 2947.

\bibitem{Bowler:ir}
R.~D.~Bowler and M.~C.~Birse,
Nucl.\ Phys.\ A {\bf 582} (1995) 655

\bibitem{Fujikawa:1979ay}
K.~Fujikawa,
Phys.\ Rev.\ Lett.\  {\bf 42} (1979) 1195.

\bibitem{Fujikawa:1980eg}
K.~Fujikawa,
Phys.\ Rev.\ D {\bf 21} (1980) 2848
[Erratum-ibid.\ D {\bf 22} (1980) 1499].

\bibitem{Salcedo:1994qy}
L.~L.~Salcedo and E.~Ruiz Arriola,
Annals Phys.\  {\bf 250} (1996) 1


\bibitem{RuizArriola:1998zi}
E.~Ruiz Arriola and L.~L.~Salcedo,
Phys.\ Lett.\ B {\bf 450} (1999) 225


\bibitem{RuizArriola:1999us}
E.~Ruiz Arriola and L.~L.~Salcedo,
arXiv:hep-th/9910230.

\bibitem{We-grav} 
S.~Weinberg, {\it Gravitation and Cosmology}, John-Wiley and Sons
(1972). 
 
\bibitem{Pelaez:2003dy}
J.~R.~Pelaez,
arXiv:hep-ph/0309292.

\bibitem{Bijnens:1996wm}
J.~Bijnens and P.~Talavera,
Nucl.\ Phys.\ B {\bf 489}, 387 (1997)


\bibitem{Toublan:1995bk}
D.~Toublan,
Phys.\ Rev.\ D {\bf 53}, 6602 (1996)
[Erratum-ibid.\ D {\bf 57}, 4495 (1998)]

\bibitem{Peris:1998nj}
S.~Peris, M.~Perrottet and E.~de Rafael,
JHEP {\bf 9805} (1998) 011


\bibitem{Bijnens:2003rc}
J.~Bijnens, E.~Gamiz, E.~Lipartia and J.~Prades,
JHEP {\bf 0304}, 055 (2003)


\bibitem{WBcoim}
W.~Broniowski, proc. of {\em Hadron Physics: Effective theories
of low-energy QCD}, Coimbra, Portugal, September 1999, AIP Conference
Proceedings {\bf 508} (1999) 380, eds. A. H. Blin and B. Hiller and M. C.
Ruivo and C. A. Sousa and E. van Beveren, AIP, Melville, New York,
hep-ph/9911204. 



\bibitem{Dorokhov:2003kf}
A.~E.~Dorokhov and W.~Broniowski,
Eur.\ Phys.\ J.\ C {\bf 32}, 79 (2003)
[arXiv:hep-ph/0305037].

\bibitem{Gilman:1967qs}
F.~J.~Gilman and H.~Harari,
Phys.\ Rev.\  {\bf 165}, 1803 (1968).

\bibitem{Weinberg:xn}
S.~Weinberg,
Phys.\ Rev.\ Lett.\  {\bf 65} (1990) 1177.

\bibitem{Svec:1996xp}
M.~Svec,
Phys.\ Rev.\ D {\bf 55}, 5727 (1997)


\bibitem{Svec:2002bz}
M.~Svec,
hep-ph/0209323.

                      
\bibitem{Svec:2002rw} M.~Svec, 
hep-ph/0210249.

\bibitem{Kaminski:2000bd}
R.~Kaminski, L.~Lesniak and K.~Rybicki,
Acta Phys.\ Polon.\ B {\bf 31}, 2265 (2000)

\bibitem{Ahmedov:2003sb}
A.~I.~Ahmedov, V.~V.~Bytev and E.~A.~Kuraev,
arXiv:hep-ph/0310032.

\bibitem{Birrell:ix}
N.~D.~Birrell and P.~C.~W.~Davies,
{\it Quantum Fields In Curved Space}, Cambridge Monographs on
Mathematical Physics, (1994) 

\bibitem{IZ80} C. Itzykson and J. B. Zuber, {\it Quantum Field
Theory}, McGraw-Hill, 1980. 

\bibitem{Luscher:1982wf}
M.~Luscher,
Annals Phys.\  {\bf 142} (1982) 359.

\bibitem{Vassilevich:2003xt}
D.~V.~Vassilevich,
Phys.\ Rept.\  {\bf 388}, 279 (2003)


\bibitem{Parker:dj}
L.~Parker and D.~J.~Toms,
Phys.\ Rev.\ D {\bf 31} (1985) 953.

\bibitem{Chan:1986jq}
L.~H.~Chan,
Phys.\ Rev.\ Lett.\  {\bf 57} (1986) 1199.






\end{thebibliography}
\end{document}